\documentclass[twocolumn]{aastex62}
\usepackage[T1]{fontenc}

\usepackage{placeins}
\usepackage{amsmath}
\usepackage{color}
\usepackage{enumerate}

\begin{document}

\newcommand\be{\begin{equation}}
\newcommand\en{\end{equation}}

\shorttitle{Observational Signatures of Buoyancy Resonances} 
\shortauthors{Bae et al.}

\title{OBSERVATIONAL SIGNATURES OF TIGHTLY WOUND SPIRALS DRIVEN BY BUOYANCY RESONANCES IN PROTOPLANETARY DISKS}

\correspondingauthor{Jaehan Bae}
\email{jbae@carnegiescience.edu}

\author[0000-0001-7258-770X]{Jaehan Bae}
\altaffiliation{NASA Hubble Fellowship Program Sagan Fellow}
\affil{Earth and Planets Laboratory, Carnegie Institution for Science, 5241 Broad Branch Road NW, Washington, DC 20015, USA}

\author[0000-0002-0786-7307]{Richard Teague}
\affil{Center for Astrophysics | Harvard \& Smithsonian, 60 Garden Street, Cambridge, MA 02138, USA}

\author[0000-0003-3616-6822]{Zhaohuan Zhu}
\affil{Department of Physics and Astronomy, University of Nevada, Las Vegas, 4505 South Maryland Parkway, Las Vegas, NV 89154, USA}

\begin{abstract}

Besides the spirals induced by the Lindblad resonances, planets can generate a family of tightly wound spirals through buoyancy resonances. The excitation of buoyancy resonances depends on the thermal relaxation timescale of the gas. By computing timescales of various processes associated with thermal relaxation, namely, radiation, diffusion, and gas-dust collision, we show that the thermal relaxation in protoplanetary disks' surface layers ($Z/R\gtrsim0.1$) and outer disks ($R\gtrsim100$~au) is limited by infrequent gas-dust collisions. The use of isothermal equation of state or rapid cooling, common in protoplanetary disk simulations, is therefore not justified. Using three-dimensional hydrodynamic simulations, we show that the collision-limited slow thermal relaxation provides favorable conditions for buoyancy resonances to develop. Buoyancy resonances produce predominantly vertical motions, whose magnitude at the $^{12}$CO emission surface is of order of $100~{\rm m~s}^{-1}$ for Jovian-mass planets, sufficiently large to detect using molecular line observations with ALMA. We generate synthetic observations and describe characteristic features of buoyancy resonances in Keplerian-subtracted moment maps and velocity channel maps. Based on the morphology and magnitude of the perturbation, we propose that the tightly wound spirals observed in TW~Hya could be driven by a (sub-)Jovian-mass planet at 90~au. We discuss how non-Keplerian motions driven by buoyancy resonances can be distinguished from those driven by other origins. We argue that observations of multiple lines tracing different heights, with sufficiently high spatial/spectral resolution and sensitivity to separate the emission arising from the near and far sides of the disk, will help constrain the origin of non-Keplerian motions. 

\end{abstract}


\section{INTRODUCTION}
\label{sec:introduction}

Recent observations have revealed a plethora of substructures in protoplanetary disks \citep[e.g.,][]{alma15,andrews18,avenhaus18}. Spiral arms, along with concentric rings and gaps, are found to be one of the most common types of substructures. They are often interpreted as an outcome of the gravitational interaction between the disk and embedded planets therein \citep[e.g.,][]{bae16c}, although we cannot rule out other possibilities, such as gravitational instabilities \citep[e.g.,][]{meru17,hall18}, stellar flybys \citep[e.g.,][]{cuello19,cuello20}, or infalling materials \citep[e.g.,][]{lesur15} until we directly detect companions.

So far, most of the spiral arms are detected in optical/near-infrared scattered light or (sub-)millimeter continuum observations observations \citep[e.g.,][]{hashimoto11,muto12,grady13,benisty15,perez16,benisty17,kraus17,andrews18,canovas18,huang18b,reggiani18,uyama18,gratton19,monnier19,keppler20,muro-arena20}. Spirals in these observations could be results of the increase in the density of emitting materials (i.e., dust grains), the increase of the temperature at the shock front, and/or the increase in the height of the scattering surface. 

With the unprecedented high spatial/spectral resolution and sensitivity the Atacama Large Millimeter/submillimeter Array (ALMA) offers, it is now possible to use molecular line observations to probe kinematics associated with spiral arms \citep[e.g.,][]{christiaens14,tang17,teague19,huang20,phuong20}, which can help better understand the origin of the spirals. It is also worth mentioning so-called velocity kinks \citep{pinte18,pinte19,pinte20} and Doppler flips \citep{sperez18,casassus19,perez20} seen in ALMA molecular line observations. While these are localized features rather than large-scale spirals, they are generally interpreted as the velocity perturbations associated with planet-induced spirals. 

Using ALMA $^{12}$CO line observations, \citet{teague19} reported three spiral arms in the velocity and temperature space in the TW~Hya disk. One of the interesting features about the spirals is that the pitch angle is very small, decreasing from 9 to 3 degrees between 70 and 200~au. This tightly wound morphology distinguishes themselves from a large number of spirals having pitch angles of 10 to 30 degrees in other protoplanetary disks \citep[e.g.,][]{huang18b,reggiani18,uyama18,monnier19,yu19}, bringing into question their origin. Because Lindblad spirals' pitch angle decreases as a function of the distance from the planet (e.g. \citealt{zhu15,baezhu18a}), it is not completely impossible to explain such a tightly wound morphology with the traditional view of Lindblad resonance-driven spirals \citep{goldreich79,goldreich80,ogilvie02,baezhu18a,baezhu18b}. One may argue that a small pitch angle could be reconciled if the planet is located sufficiently far inward of the observed spirals. In this case, however, it is unclear why we do not observe spirals near the planet but only far from it because we expect spirals generate stronger perturbations closer to the planet. 

\subsection{Theoretical Background: Buoyancy Resonances}

Here we examine an alternative: buoyancy resonances \citep{zhu12,lubow14,McNally2020}. Let us consider a gas parcel that is vertically displaced from the equilibrium position. Its response to the perturbation can be described with the vertical buoyancy frequency (a.k.a. Brunt-B\"ais\"al\"a frequency) 
\be
\label{eqn:buoyancy_frequency}
N_Z^2 = {g \over \gamma} {\partial \over \partial Z} \left[  \ln \left( P \over \rho^\gamma \right) \right ],
\en
or
\be
\label{eqn:buoyancy_frequency_ideal}
N_Z^2 = {g \over \gamma} {\partial \over \partial Z} \left[  \ln \left(T \over \rho^{\gamma-1} \right) \right ]
\en 
adopting the ideal gas law. In the above equations, $g$ is the gravitational acceleration, $\gamma$ is the adiabatic index, $P$ is the gas pressure, $\rho$ is the gas density, and $T$ is the gas temperature. When the disk temperature is dominated by the stellar irradiation, the disk is hotter near the surface and colder near the midplane \citep{chiang97,dalessio98,dartois03,rosenfeld13}. With a positive vertical temperature gradient and $\gamma \geq 1$, Equation (\ref{eqn:buoyancy_frequency_ideal}) yields $N_Z^2 > 0$. We thus expect the gas parcel to vertically oscillate around its equilibrium position with a frequency $N_Z$. 

When the buoyancy frequency matches with the forcing frequency which, in this case, is planet's orbital frequency, buoyancy resonances can develop \citep{zhu12,lubow14}. Buoyancy resonances give a rise to the density and velocity perturbations along a family of trailing spirals. As we will show below, one of the main characteristics of buoyancy spirals is that they are very tightly wound compared with Lindblad spirals, in particular in the vicinity of the planet.

It is worth pointing out that the thermodynamic properties of the disk is important in the development of buoyancy resonances. In order for buoyancy resonances to fully develop, the timescale for the gas to respond to thermal perturbations (hereafter relaxation timescale $t_{\rm relax}$) has to be longer than the timescale associated with the buoyancy: $t_{\rm relax} \gtrsim N_Z^{-1}$. When $t_{\rm relax} \ll N_Z^{-1}$, the gas behaves isothermally and buoyancy resonances are expected to be weak or absent. 

In this paper, we show planets can excite spirals through buoyancy resonances, which can be detectable using molecular line observations with ALMA. The paper is organized as follows. In Section \ref{sec:thermal_relaxation}, we describe three processes that determine the relaxation timescale of the gas -- radiation, diffusion, and gas-dust collision -- and compute the corresponding timescales using a TW~Hya disk model. We show that thermal relaxation in the surface layers ($Z/R \gtrsim 0.1$) and outer regions ($R \gtrsim 100$~au) of protoplanetary disks can be limited by insufficient gas-dust collision. In Section \ref{sec:simulation}, we present three-dimensional hydrodynamic simulations and show that the collision-limited slow thermal relaxation provides favorable conditions for buoyancy resonances. In Section \ref{sec:simobs}, we generate synthetic observations and show that the spirals driven by buoyancy resonances are observable with molecular line observations using ALMA. Based on the tightly wound morphology and the magnitude of velocity perturbations, we propose that the velocity spiral seen in TW~Hya could be driven by a (sub-)Jovian-mass planet at 90~au. In Section \ref{sec:discussion}, we show that the relaxation timescale is comparable to or longer than the dynamical timescale under a broad range of conditions, and discuss its implications for the development of buoyancy resonances, planet-induced gap profiles, and hydrodynamic instabilities. We also discuss potential ways to discriminate non-Keplerian motions driven by buoyancy resonances from those driven by other mechanisms, including Lindblad resonance, corrugated vertical flows, and gas pressure changes. We summarize our findings and conclude in Section \ref{sec:summary}.

\section{Thermal Relaxation of the Disk Gas}
\label{sec:thermal_relaxation}

\subsection{Disk Model}
\label{sec:disk_model}

As one of the motivations of this work is to explain the tightly wound spirals observed in the TW~Hya disk \citep{teague19}, we adopt the disk density and temperature profiles similar to the one constrained for the disk in \citet{huang18a}.

The gas surface density follows
\be
\label{eqn:sigma}
\Sigma_{\rm g}(R)   =  \Sigma_p \left({R/ R_p}\right)^{-p}.
\en
where $\Sigma_p$ is the surface density at $R_p=90$~au and $p=0.9$. We choose $\Sigma_p = 3.8~{\rm g~cm}^{-2}$ such that the total disk gas mass within 210~au is 0.05~$M_\odot$, broadly consistent with observational estimates for the TW Hya disk \citep{bergin13}. Throughout this paper we use $R=r\sin{\theta}$ for the cylindrical radius and $Z=r\cos{\theta}$ for the height, where $r$ and $\theta$ are the radius and the polar angle in the spherical coordinates, respectively.

Following the prescription in \citet{dartois03}, the gas temperature is parameterized as follows.
\begin{eqnarray}
T_{\rm g}(Z)
=\begin{cases}
T_{\rm atm} + (T_{\rm mid} - T_{\rm atm}) \cos^4\left( {\pi \over 2} {Z \over Z_q} \right)&
\text{if}~Z < Z_q \\
T_{\rm atm}&
\text{if}~Z \geq Z_q \\
\end{cases}
\label{eqn:temperature}
\end{eqnarray}
Here, the midplane and atmosphere temperatures are a function of the cylindrical radius $R$ following
\be
T_{\rm atm}(R)   =  T_{\rm atm,p} \left({R/R_p}\right)^{-q}
\label{eqn:temperature_atm}
\en
and
\be
T_{\rm mid}(R)   =  T_{\rm mid,p} \left({R/R_p}\right)^{-q},
\label{eqn:temperature_mid}
\en
where $T_{\rm atm,p} = 44.5$~K, $T_{\rm mid,p} = 14.2$~K, and $q=0.47$ \citep{zhang17,huang18a}. In Equation (\ref{eqn:temperature}), $Z_q (R) = 4H_{\rm mid} (R)$ where $H_{\rm mid}$ is the gas scale height determined with the disk midplane temperature. In terms of the disk aspect ratio, the above temperature profile at the midplane corresponds to $H_{\rm mid}/R = 0.075\times(R/R_p)^{0.265}$, assuming $0.88~M_\odot$ for the stellar mass \citep{andrews12,huang18a} and 2.4 for the mean molecular weight of the gas.

Using the above surface density and temperature profiles, we construct the three-dimensional gas density distribution that satisfies the vertical hydrostatic equilibrium: 
\begin{equation}
- {GM_* Z \over (R^2 + Z^2)^{3/2}} - {1 \over \rho_{\rm g}}{\partial P \over \partial Z} = 0.
\end{equation}
Solving the above equation results in the vertical density distribution that follows
\begin{equation}
\rho_{\rm g}(Z) = \rho_{\rm g, mid} {c_{s,{\rm mid}}^2 \over c_s^2(Z)} \exp \left[ - \int_{0}^{Z} {{1\over c_s^2(Z')}  {GM_*Z' \over (R^2 + Z'^2)^{3/2}}} {\rm d}Z' \right],
\end{equation}
where $c_{s,{\rm mid}}$ and $c_s (Z)$ denote the sound speed at the disk midplane and at height $Z$.
We then compute the angular velocity $\Omega$ that satisfies the radial force balance, taking into account the gas pressure gradient:
\begin{equation}
\Omega^2 = \Omega_K^2 \sin\theta + {1 \over \rho_{\rm g} r \sin^2\theta} {\partial P \over \partial r}.
\end{equation}
Here, $\Omega_K = \sqrt{GM_*/R^3}$ is the Keplerian angular velocity.
The initial radial and meridional velocities are set to zero.

\subsection{Thermal Relaxation Timescale}
\label{sec:thermodynamics}

\begin{figure}[t!]
    \centering
    \includegraphics[width=0.48\textwidth]{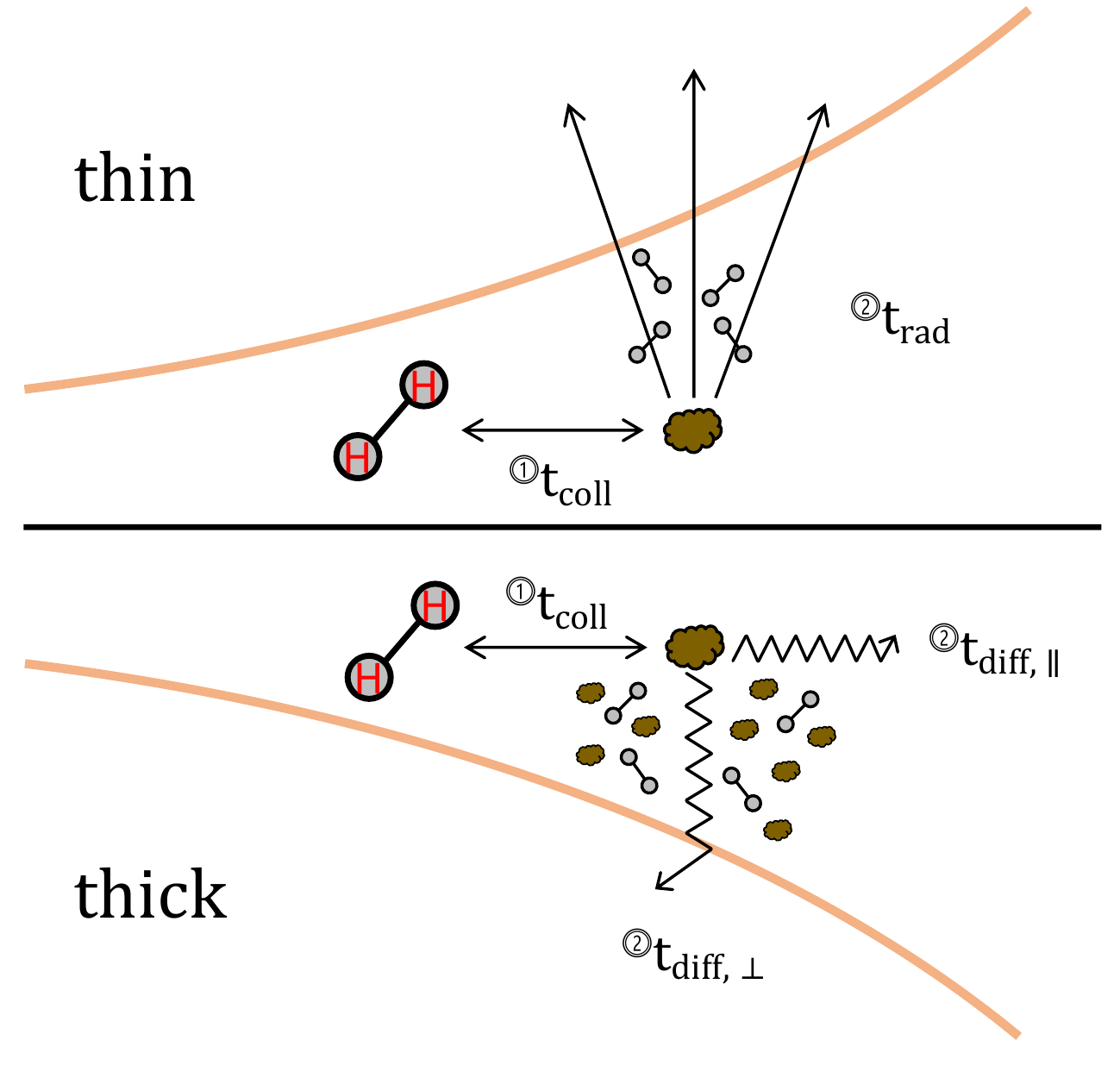}
    \caption{A schematic diagram showing the cooling process of the gas in protoplanetary disks. The cooling of the gas is a sequential, two-step process. Hydrogen molecules -- the most abundant but poorly emitting species -- first have to collide with dust grains to lose their energy. Dust grains subsequently radiate away the excess thermal energy. The thermal photons from dust grains escape the disk freely when the disk is optically thin (the upper half of the diagram), while they do so after multiple re-absorption/emission (i.e., diffusion) when the disk is optically thick (the lower half of the diagram).}
    \label{fig:cooling}
\end{figure}

Protoplanetary disks are a mixture of gas and dust. Hydrogen molecules dominate the total mass and thermal energy, but they are inefficient at emitting radiation. Let us consider a situation where a gas parcel has been perturbed from its equilibrium state to have a higher temperature. In order for the gas to lose its thermal energy, hydrogen molecules first have to transfer their kinetic energy to the surrounding dust grains. Dust grains then radiate away the excess thermal energy. This process can thus be understood as a sequential, two-step process (see Figure \ref{fig:cooling} for a schematic diagram).

When the collision between hydrogen molecules and dust grains is sufficiently frequent, the gas cools at the rate dust grains radiate away their thermal energy; in this case, gas and dust are thermally coupled. In the optically thin regime, thermal photons emitted by dust grains can freely escape from the disk. The relaxation timescale of the gas can be described by the radiative timescale of dust grains
\be
\label{eqn:tcool_rad}
t_{\rm rad} = {c_V \over 16 \kappa_P \sigma_{\rm SB} T_{\rm g}^3},
\en
where $c_V$ denotes the specific heat capacity of the gas, $\kappa_P$ is the Planck mean opacity of the dust, and $\sigma_{\rm SB}$ is the Stefan-Boltzmann constant.

In the optically thick regime, thermal photons emitted from dust grains are absorbed and emitted by other grains multiple times before they eventually escape the disk. In this case, the cooling timescale can be characterized by the diffusion of photons. The diffusion timescale associated with the length scale $\lambda_{\rm diff}$ can be written as
\be
\label{eqn:tcool_diff}
t_{\rm diff} =  {\lambda_{\rm diff}^2 \over \langle D \rangle},
\en
where $D$ is the diffusion coefficient defined as
\be
\label{eqn:diff_coeff}
D = {16 \sigma_{\rm SB} T_{\rm g\\}^3 \over 3 c_V  \kappa_R \rho_{\rm g}^2 }
\en
with $\kappa_R$ being the Rosseland mean opacity of the dust and 
the brackets $\langle~\rangle$ denoting an average over $\lambda_{\rm diff}$ \citep{malygin17}. While the density, temperature, and opacity are expected to vary moderately along the radial and azimuthal directions, their variation can be  larger along the vertical direction due to the vertical stratification of the disk. We thus define in-plane diffusion timescale $t_{\rm diff, \parallel}$ and vertical diffusion timescale $t_{\rm diff, \perp}$, for which the diffusion coefficient is averaged along the radial and vertical directions, respectively. The overall diffusion timescale considering both in-plane and vertical diffusion can be estimated as 
\be
\label{eqn:tcool_diff2}
t_{\rm diff} = \left( {1 \over t_{\rm diff,\parallel}} + {1 \over t_{\rm diff,\perp}} \right)^{-1}.
\en

For a given length scale $\lambda_{\rm diff}$, we find that $t_{\rm diff, \parallel}$ and $t_{\rm diff, \perp}$ are comparable when $\lambda_{\rm diff} \ll H_{\rm g}$, not surprisingly, where $H_{\rm g}$ is the gas scale height. However, one can be larger than the other by a factor of a few when $\lambda_{\rm diff} \gtrsim H_{\rm g}$. It is also worth mentioning that it is possible that the length scale of the temperature perturbation along the radial/azimuthal direction and that along the vertical direction differ (see e.g., \citealt{miranda20b}).

We note that, with the assumption that the disk has a constant temperature vertically, along with $\tau_R \equiv \int {\kappa_R \rho_{\rm g}} {\rm d}Z$ and $\Sigma_{\rm g} \simeq \rho_{\rm g} H_{\rm g}$, combining Equations (\ref{eqn:tcool_diff}) and (\ref{eqn:diff_coeff}) approximates to 
\be
t_{\rm diff, \perp} \simeq {3c_V\Sigma_{\rm g} \over 16\sigma_{\rm SB} T_{\rm g}^3} \tau_R, 
\en
a formula that is often adopted to account for optically thick cooling in vertically-integrated, one-/two-dimensional framework.

Let us now move on to the situation where the gas-dust collision is not sufficiently frequent. In this case, cooling of the gas is limited by the gas-dust collision rate (assuming lack of other cooling mechanisms, such as molecular/atomic line cooling; see below). Since the assumption of thermal equilibrium is no longer valid, we define the dust temperature $T_{\rm d}$. The gas-dust collisional timescale $t_{\rm coll}$ can be written as
\be
\label{eqn:tcool_coll}
t_{\rm coll} = {{\rho_{\rm g} c_V} \over {\Lambda_{\rm coll}}} | T_{\rm g} - T_{\rm d}|,
\en
where $\Lambda_{\rm coll}$ is the cooling rate per unit volume via gas-dust collisions for which we follow \citet{burke83}:
\be{}
\label{eqn:coll_rate}
\Lambda_{\rm coll} = \int_{a_{\rm min}}^{a_{\rm max}} 2k_B | T_{\rm g} - T_{\rm d} | \bar{\alpha_T} n_{\rm g} n_{\rm d}(a) \sigma_{\rm d}(a) v_{\rm th} {\rm d}a.
\en
In the above equation, $a_{\rm min}$ and $a_{\rm max}$ denote the minimum and maximum dust grain sizes, $k_B$ is the Boltzmann constant, $\bar{\alpha_T} = 0.5$ is the thermal accommodation coefficient that characterizes the efficiency of the heat transfer between gas molecules and dust grains, $n_{\rm g}$ is the number density of gas molecules, $n_{\rm d}(a)$ is the number density of dust particles with size $a$, $\sigma_{\rm d} = \pi a^2$ is the geometrical cross section of dust particles, and $v_{\rm th} = (8k_BT_{\rm g}/\pi m_g)^{1/2}$ is the thermal velocity of the gas. Using second and third moment of the dust grain size, which are defined as
\be
\langle a^2 \rangle = \int_{a_{\rm min}}^{a_{\rm max}} {a^2 n_{\rm d}(a)} \rm{d}a
\en
and
\be
\langle a^3 \rangle = \int_{a_{\rm min}}^{a_{\rm max}} {a^3 n_{\rm d}(a)} \rm{d}a,
\en
Equation (\ref{eqn:coll_rate}) can be written as 
\be
\label{eqn:coll_rate2}
\Lambda_{\rm coll} =  2 k_B | T_{\rm g} - T_{\rm d} | \bar{\alpha_T} \left( {3 \over 4 \rho_s} \right) \left( {\langle a^2 \rangle \over \langle a^3 \rangle} \right) \left( {\rho_{\rm g}^2 v_{\rm th} \over m_{\rm g}} \right) \left( {\rho_{\rm d} \over \rho_{\rm g}} \right),
\en
where $\rho_s$ is the bulk density of dust grains while $\rho_{\rm d}$ is the mass density of dust grains. Then, combining Equations (\ref{eqn:tcool_coll}) and (\ref{eqn:coll_rate2}), the collisional timescale can be written as 
\be
\label{eqn:tcool_coll2}
t_{\rm coll} = {c_V \over {2 k_B \bar{\alpha_T}}} \left( {4 \over 3} \rho_s \right) \left( {\langle a^3 \rangle \over \langle a^2 \rangle} \right) \left( {m_{\rm g} \over \rho_{\rm g} v_{\rm th}} \right) \left( {\rho_{\rm d} \over \rho_{\rm g}} \right)^{-1}.
\en
Note that the collisional timescale depends on the mean dust grain size and the local dust-to-gas mass ratio, which are dependent upon the grain size distribution and the level of turbulence of the disk among many others.

Then, taking into account radiation, diffusion, and gas-dust collisional timescales, the relaxation timescale of the gas can be written as 
\be
\label{eqn:trelax}
t_{\rm relax} = t_{\rm coll} + \max(t_{\rm rad}, t_{\rm diff}).
\en
Again, this equation shows that the collision between gas and dust has to precede thermal emission of dust grains. When the gas-dust collision is not sufficiently frequent, the collisional energy exchange can be the bottleneck of the cooling process and thus the overall cooling timescale is determined by the collisional timescale (i.e., $t_{\rm relax} \simeq t_{\rm coll}$). When the gas-dust collision is frequent, the gas and dust are in thermal equilibrium and the gas cools over the timescale $t_{\rm rad}$ or $t_{\rm diff}$ depending on the optically thickness of the disk (i.e., $t_{\rm relax} \simeq {\rm max}(t_{\rm rad}, t_{\rm diff})$). The transition between the optically thick and thin regimes occur when $t_{\rm diff} \simeq t_{\rm rad}$ \citep{malygin17}. 

In order to make a quantitative comparison between $t_{\rm rad}$, $t_{\rm coll}$,  and $t_{\rm diff}$, we compute the timescales using the disk model described in Section \ref{sec:disk_model}. To do so, we first need to define the spatial/size distribution of dust grains as it dictates the gas-dust collision rate but also the radiative/diffusion timescales through the opacity.  We adopt a maximum grain size that is decreasing over radius: $a_{\rm max}(R) = 1~{\rm mm}~(R/30~{\rm au})^{-2}$. This choice is  motivated by the fact that the (sub-)millimeter continuum emission of the TW~Hya disk is confined within about 60~au \citep[e.g.,][]{andrews16,tsukagoshi16}. With the vertically-integrated total dust-to-gas mass ratio fixed to 0.01 at each radius (i.e., $\Sigma_{\rm d}/\Sigma_{\rm g} = 0.01$), we distribute the dust mass between $0.1~\mu$m and $a_{\rm max}$ adopting a power-law dust size distribution with a power-law index $-3.5$: $n_{\rm d}(a) \propto a^{-3.5}$. We then determine the vertical scale height of each dust species assuming that vertical settling is balanced by turbulence mixing characterized by $\alpha=10^{-3}$. This results in a dust scale height of
\be
\label{eqn:scale_dust}
H_{\rm d} (R, a) = H_{\rm g} (R) \times \min \left( 1, \sqrt{{\alpha \over {\min({\rm St}(a), 1/2)(1+{\rm St}(a)^2)}}} \right) 
\en
\citep{dullemond04,birnstiel10}, where St$(a)\equiv (\rho_{\rm s}a/\rho_{\rm g}v_{\rm th})\Omega_K$ is the Stokes number of particle having size $a$ and we assume a grain internal density $\rho_s = 1.67~{\rm g~cm}^{-3}$ (see below). Then, the vertical density distribution of each dust species is obtained following
\be
\rho_d(R,Z,a) = {\Sigma_d(R,a) \over \sqrt{2\pi}H_d(R,a)} \exp\left( - {Z^2 \over 2H_d(R,a)^2} \right).
\en

\begin{figure*}[t!]
    \centering
    \includegraphics[width=0.95\textwidth]{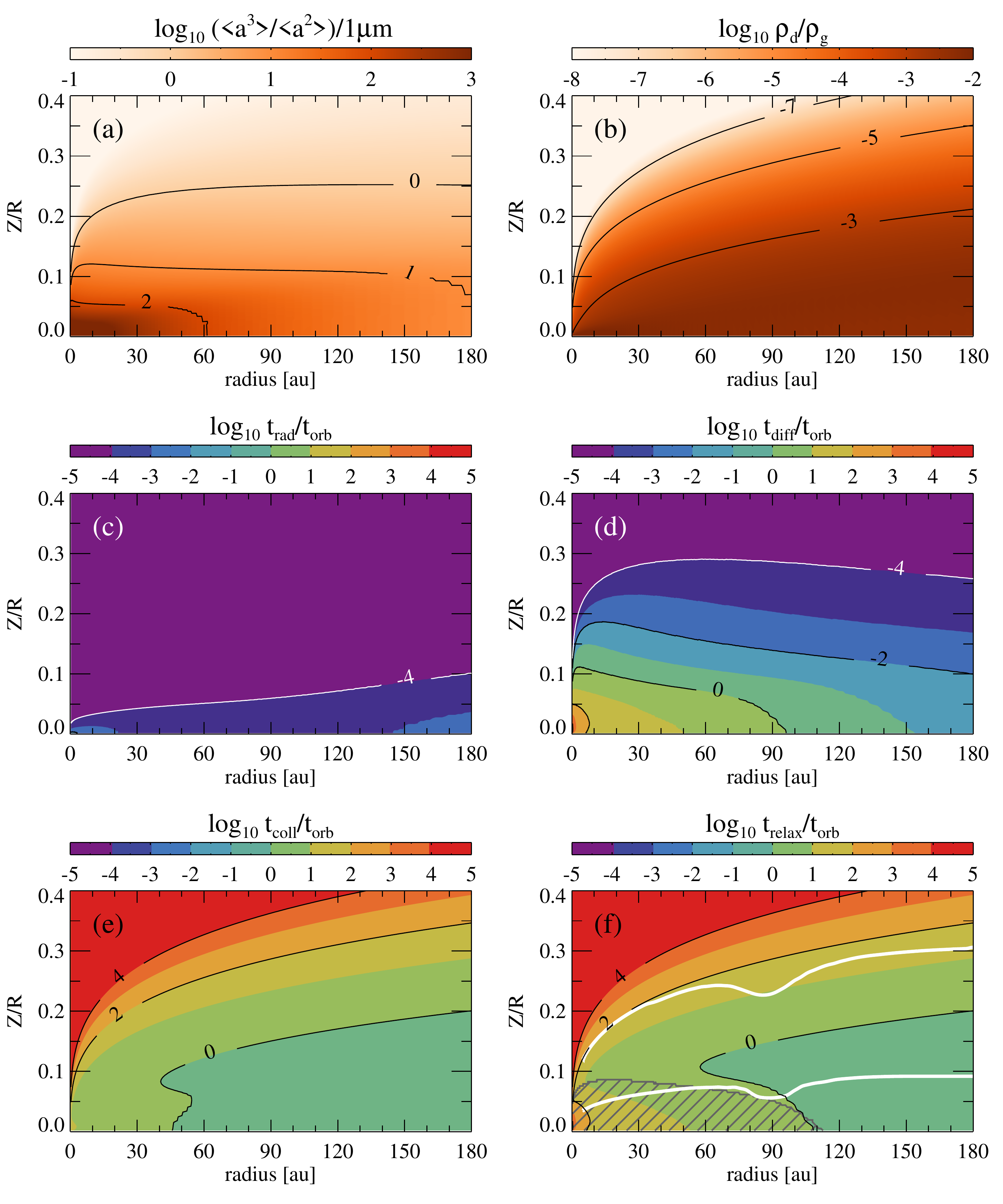}
    \caption{(a) The mean dust grain size $\bar{a} \equiv \langle a^3 \rangle/\langle a^2 \rangle$. (b) The dust-to-gas mass ratio $\rho_d/\rho_g$. The (c) radiative, (d) diffusion, and (e) collisional timescales in units of the local orbital time. (f) The  relaxation timescale $t_{\rm relax}$, defined as in Equation (\ref{eqn:trelax}). In panel (f), the shaded region in gray shows where $t_{\rm diff} \geq t_{\rm coll}$. We emphasize that the thermal relaxation of the gas is limited by infrequent gas-dust collision in the most region of the disk ($Z/R \gtrsim 0.1$) and $t_{\rm relax}$ is comparable to or longer than the orbital timescale. The white curves in panel (f) show the emission surfaces of (upper) $^{12}$CO~$J=3-2$ and (lower) $^{13}$CO~$J=3-2$ lines from the synthetic observation presented in Section \ref{sec:simobs}, based on the standard model with a $0.5~M_{\rm Jup}$ planet.}
    \label{fig:cooling_time}
\end{figure*}

The resulting mean dust particle size $\langle a^3 \rangle/\langle a^2 \rangle$ and dust-to-gas mass ratio $\rho_d/\rho_g$ are shown in Figure \ref{fig:cooling_time}a and b. Larger grains settle near the midplane due to shorter settling times. Since larger grains contain a larger fraction of the total dust mass, the dust-to-gas mass ratio decreases over height. Dust grains with sizes $\gtrsim 100~\mu$m, which are believed to dominate the (sub-)millimeter continuum emission, are confined in radius within the inner $\sim60$~au, while $\mu$m-sized grains extend much further out, consistent with both (sub-)millimeter and optical/near-infrared observations of TW~Hya \citep[e.g.,][]{andrews16,tsukagoshi16,debes13,debes17,vanboekel17}.

\begin{figure*}[t!]
    \centering
    \includegraphics[width=\textwidth]{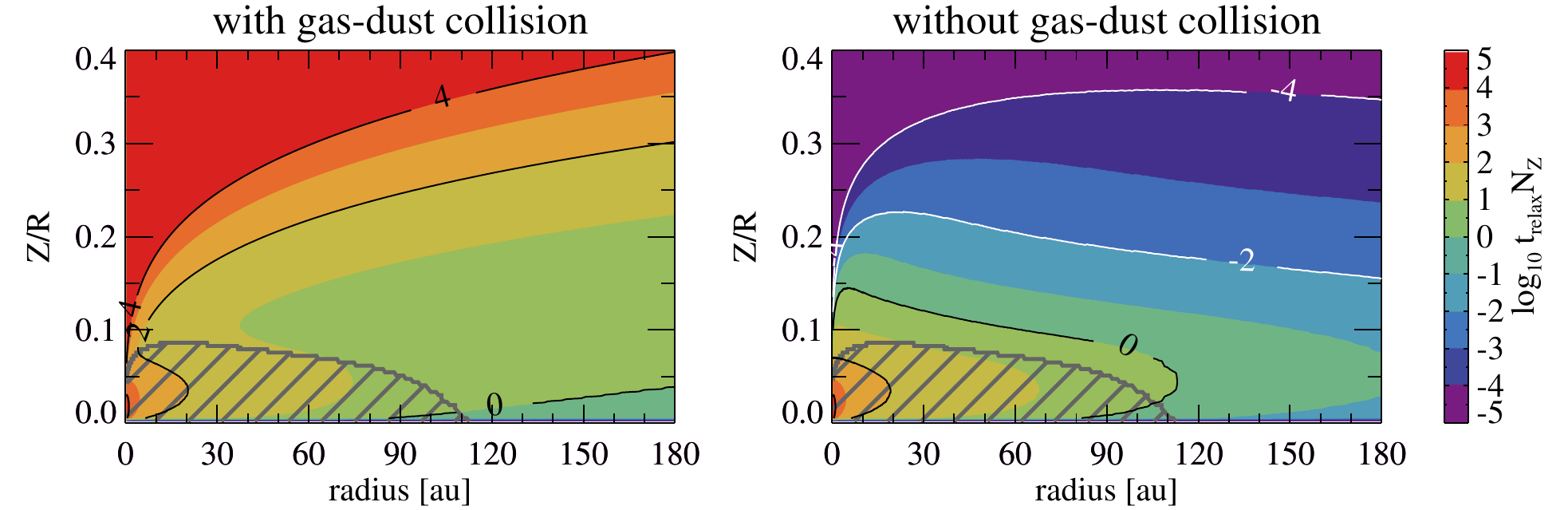}
    \caption{$t_{\rm relax} N_Z$ in a logarithmic scale (left) considering gas-dust collision and (right) without considering gas-dust collision. Note that when gas-dust collision is considered $t_{\rm relax} \gtrsim N_Z^{-1}$ in the entire disk except near the midplane, providing favorable conditions for buoyancy resonances to develop. The shaded region in gray shows where $t_{\rm diff} \geq t_{\rm coll}$.}
    \label{fig:t_relax_Nz}
\end{figure*}

Next, we calculate the dust opacity in each grid cell based on the dust distribution obtained as above, adopting the opacity model from the DSHARP collaboration \citep{birnstiel18}. In this opacity model, grains are assumed as a mixture of  water ice, astronomical silicates, troilite, and refractory organic material, having a bulk density of $1.67~{\rm g~cm}^{-3}$. The optical constants to compute the DSHARP opacity are originally from \citet{henning96,draine03,warren08}.
The absorption and scattering opacities are calculated following
\be
\kappa_{\nu}^{\rm abs,sca} (R,Z) = {{ {\sum_{i} \kappa_\nu^{\rm abs,sca} (a_i) \rho_{\rm d} (R,Z,a_i)}} \over {\sum_{i} \rho_{\rm d} (R,Z,a_i)}},
\en
where the subscription $\nu$ shows the frequency dependency of the opacity.
The Rosseland and Planck mean opacities are calculated as
\be
\kappa_R (R, Z) = \left[ {{\int_{0}^{\infty} {{ 1 \over \kappa_\nu^{\rm ext} (R, Z)}} {dB_\nu \over dT} d\nu} \over {\int_{0}^{\infty} {{dB_\nu \over dT} d\nu}}} \right]^{-1}
\en
and
\be
\kappa_P (R, Z) = {{\int_{0}^{\infty} {\kappa_\nu^{\rm abs} (R, Z)} B_\nu d\nu} \over {\int_{0}^{\infty} {B_\nu d\nu}}},
\en
where $\kappa_\nu^{\rm ext} = \kappa_\nu^{\rm abs} + \kappa_\nu^{\rm sca}$.

The radiation timescale $t_{\rm rad}$ calculated as in Equation (\ref{eqn:tcool_rad}) is shown in Figure \ref{fig:cooling_time}c. The radiation timescale is orders of magnitude shorter than the dynamical timescale everywhere in the disk. At a given radius, the radiation timescale decreases over height because of its steep temperature dependency ($t_{\rm rad} \propto T_{\rm g}^{-3}$).

Figure \ref{fig:cooling_time}d presents the diffusion timescale $t_{\rm diff}$.
Diffusion is slower when the optical depth is larger. So $t_{\rm diff}$ is greater at smaller radii and near the midplane, and is a decreasing function of $R$ and $Z$. In particular, note that $t_{\rm diff}$ drops exponentially over height because of the density dependency ($t_{\rm diff} \propto \rho_{\rm g}^2$). We also note that the diffusion timescale depends on the temperature perturbation length scale as $t_{\rm diff} \propto \lambda_{\rm diff}^2$. Here, we opt to use $\lambda_{\rm diff} = H_{\rm g}$. In reality, the length scale of any perturbations can range from $\ll H_{\rm g}$ to the thickness of the disk, which is a few scale heights. However, as we will show below (see also Section \ref{sec:t_relax}), the thermal relaxation in the surface layers ($Z/R \gtrsim 0.1$) is limited by infrequent gas-dust collision, insensitive to the choice of $\lambda_{\rm diff}$.

Figure \ref{fig:cooling_time}e shows the gas-dust collisional timescale $t_{\rm coll}$. For given gas density and temperature structures, the collisional timescale is set by the mean grain size and the dust-to-gas mass ratio ($t_{\rm coll} \propto \langle a^3 \rangle/ \langle a^2 \rangle,~(\rho_{\rm d}/\rho_{\rm g})^{-1}$). As the dust-to-gas mass ratio drops exponentially over height, gas molecules have significantly less frequent collisions with dust grains. This makes the collisional timescale orders of magnitude longer than the dynamical timescale in the surface layers.

The relaxation timescale $t_{\rm relax}$, calculated as in Equation (\ref{eqn:trelax}), is shown in Figure \ref{fig:cooling_time}f. The plot clearly shows that the assumption of thermal equilibrium between gas and dust is not necessarily valid in the outer and surface regions of the disk because of the long collisional timescale there. Note that this picture is in a good agreement with previous studies of thermal relaxation in protoplanetary disks \citep[e.g.,][]{malygin17,barranco18,pfeil19}. In Section \ref{sec:t_relax}, we explore how different assumptions on the diffusion length scale, level of disk turbulence, grain size distribution, disk mass, and the existence of a gap in the disk can affect the cooling timescales. As we will show, the thermal relaxation of the gas in surface layers of protoplanetary disks is limited by infrequent gas-dust collision over a broad range parameter space.

As we mentioned in Section \ref{sec:introduction}, buoyancy resonances require adiabatic responses to thermal perturbations to fully develop (i.e., $t_{\rm relax} \gtrsim N_Z^{-1}$). To see how $t_{\rm relax}$ compares with $N_Z^{-1}$, we present $t_{\rm relax} N_Z$ in Figure \ref{fig:t_relax_Nz}. As shown, when gas-dust collision is taken into account, $t_{\rm relax} \gtrsim N_Z^{-1}$ in the entire disk except at the midplane where $N_Z$ is zero due to the symmetry across the midplane. This suggests that most part of the disk, including the surface layers CO lines probe, has favorable conditions for buoyancy resonances to develop. In contrast, if gas-dust collision is neglected, the surface layers have $t_{\rm relax} \ll N_Z^{-1}$  suggesting that buoyancy resonances are weak or unlikely to develop there.

In addition to the cooling processes we considered above, atomic and molecular line cooling plays a role in the surface layers of protoplanetary disks \citep[e.g.,][]{gorti11,du14,kama16,facchini18}. The exact height beyond which line cooling dominates depends on the underlying thermal/chemical properties of the disk as well as the stellar/external irradiation. While including comprehensive thermo/photo-chemistry requires full thermochemical radiative transfer calculations, which is beyond the scope of the paper, here we test the potential effect line cooling would have on buoyancy resonances by assuming that the disk gas cools efficiently beyond a certain height of the disk $Z=Z_{\rm line}$. Taking this into account, we adopt the following form for the relaxation timescale considering diffusion, radiation, gas-dust collision, and line cooling
\be
\label{eqn:t_relax}
t_{\rm relax} = \left[ t_{\rm coll} + \max(t_{\rm rad}, t_{\rm diff}) \right] \exp\left[ -\left( {Z \over Z_{\rm line}} \right)^{-12}\right].
\en
The exponential term on the right-hand-side of the equation is added to mimic the effect of line cooling\footnote{The exponent $-12$ is chosen such that $t_{\rm relax}$ falls sufficiently rapidly over height so $t_{\rm relax} \ll t_{\rm orb}$ at the upper boundary of the simulation domain.}. 

In our fiducial models, we adopt $Z_{\rm line} = 4H_{\rm g}$ which corresponds to $Z/R = 0.3$ at the radial location of the planet in our simulations, 90~au. This choice is motivated by the thermochemical model of TW~Hya presented in \citet{kama16}, where the major atomic line emission, including [C~I], [C~II], and [O~I] lines, originates from $Z/R \sim 0.3 - 0.4$ (see their Figure D.2.). However, it is important to note that the exact line cooling rate is dependent upon various factors, including the gas phase abundance of the coolants and electron number density which is determined by UV flux as well as full chemical chains. To test the effect of line cooling, we ran additional simulations adopting $Z_{\rm line} = 2H_{\rm g}$ ($Z/R = 0.15$ at 90~au; Section \ref{sec:zline_2h}). In this model, the $^{12}$CO molecular line probes the layers where cooling is rapid, dominated by line cooling.

For the sake of reproducing the hydrodynamic simulations we will present in the following sections, we provide parameterized fits to the diffusion and collisional timescales:
\be
\label{eqn:cooling_diff}
{t_{\rm diff} \over t_{\rm orb}}  = 2.4 \left( {R \over 90~{\rm au}} \right)^{-3.9} \exp\left[ -\left( {Z \over Z_{\rm diff}} \right)^{0.9} \right],
\en
where $Z_{\rm diff} = 2.0~{\rm au}~(R/90~{\rm au})^{1.4}$,
and
\be
\label{eqn:cooling_coll}
{t_{\rm coll} \over t_{\rm orb}}  = 0.4 \left( {R \over 90~{\rm au}} \right)^{-1.1} \exp\left[ \left( {Z \over Z_{\rm coll}} \right)^{2.2} \right],
\en
where $Z_{\rm coll} = 11.7~{\rm au}~(R/90~{\rm au})^{1.2}$.

\begin{figure*}[ht!]
    \centering
    \includegraphics[width=1\textwidth]{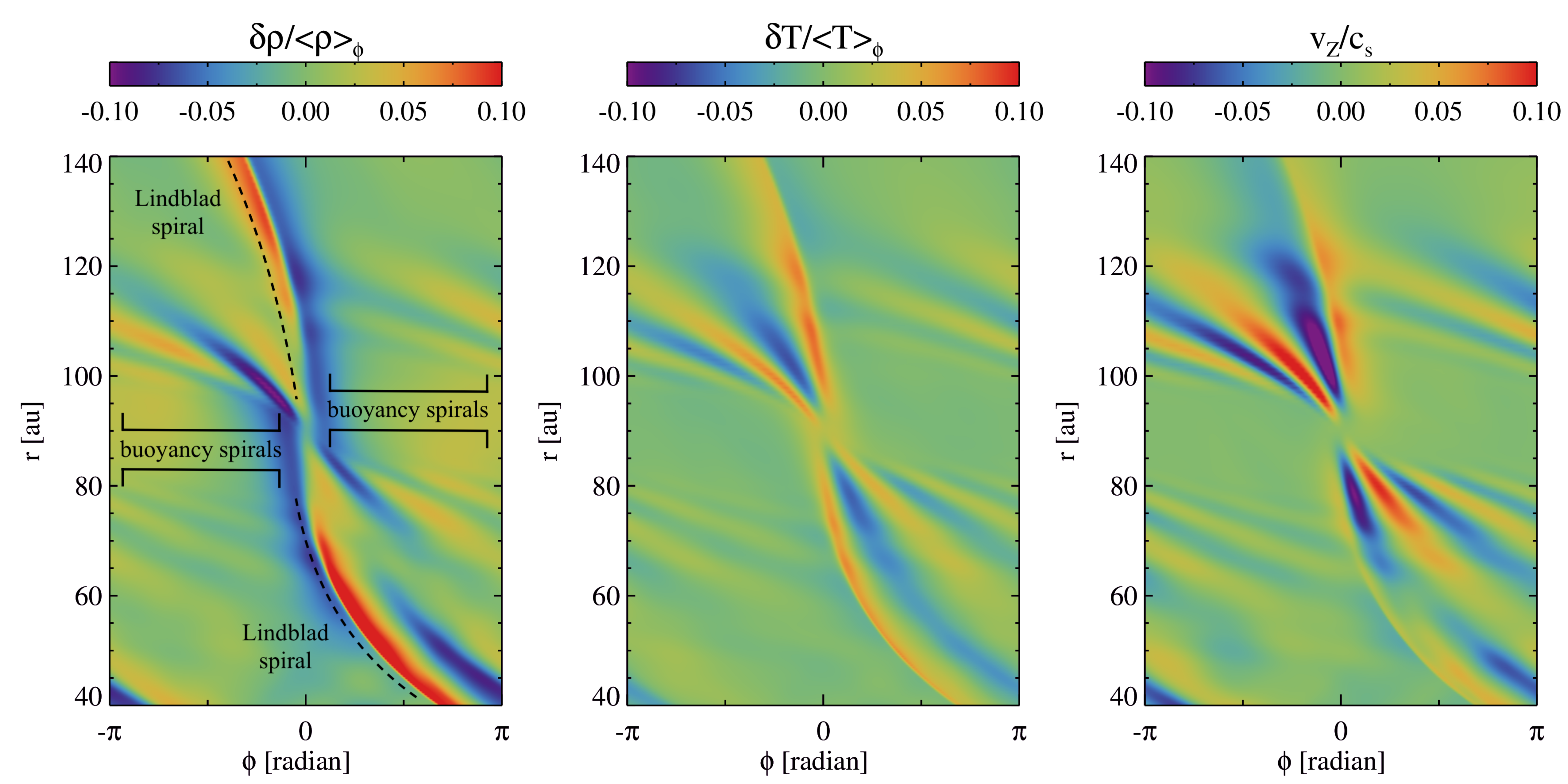}
    \caption{Results from the standard model with $M_p = 0.5~M_{\rm Jup}$ taken at $t=5~t_{\rm orb}$. From left to right, the perturbed density $\delta\rho/\langle \rho \rangle_\phi$, perturbed temperature $\delta T / \langle T \rangle_\phi$, and vertical velocity $v_Z$ in units of local sound speed at $Z/R=0.23$ ($\simeq 3~H$ at 90~au) in a $\phi-r$ plane. Here, $\langle \rangle_\phi$ denotes average over the azimuthal direction and $\delta\rho \equiv \rho - \langle \rho \rangle_\phi$ and $\delta T \equiv T - \langle T \rangle_\phi$. The planet is located in the midplane, at $(\phi, r) = (0~{\rm radian}, 90~{\rm au})$. The vertical velocity $v_Z$ is computed from radial and meridional velocities in the spherical coordinates. For a more straightforward comparison with observations later in Section \ref{sec:simobs},  motions toward the disk midplane are shown with red colors and positive numbers while motions toward the disk surface are shown with blue colors and negative numbers, such that they match with the red- and blue-shift convention.}
    \label{fig:hydro_wtcoll}
\end{figure*}

\section{Hydrodynamic Simulations}
\label{sec:simulation}

\subsection{Hydrodynamic Equations Solved}
\label{sec:hydro_equations}

We solve the hydrodynamic equations for mass, momentum, and energy conservation in the three-dimensional spherical coordinates $(r,\theta,\phi)$ using FARGO 3D \citep{benitez16,masset00}:
\be\label{eqn:mass}
{\partial \rho_{\rm g} \over \partial t} + \nabla \cdot (\rho_{\rm g} v) = 0,
\en
\be\label{eqn:momentum}
\rho_{\rm g} \left( {\partial v \over \partial t} + v \cdot \nabla v \right) = - \nabla P - \rho_{\rm g} \nabla (\Phi_* + \Phi_p) + \nabla \cdot \Pi,
\en
\be\label{eqn:energy}
{\partial e \over \partial t} + \nabla \cdot (ev) = - P \nabla \cdot v + Q_{\rm relax}.
\en
In the above equations, $\rho_{\rm g}$ is the gas density, $v$ is the velocity vector, $P$ is the gas pressure, $\Phi_* = -GM_*/r$ is the gravitational potential of the central star having mass $M_*$, $\Phi_p$ is the gravitational potential of the planet,  $\Pi$ is the viscous stress tensor, $e$ is the internal energy per unit volume, and $Q_{\rm relax}$ is the rate at which the gas thermally relaxes to the initial state (see Section \ref{sec:thermodynamics}). Note that $Q_{\rm relax}$ can be either a positive or a negative value depending on whether the gas is colder or hotter than the initial equilibrium temperature.

The gravitational potential of the planet is computed as 
\be
\label{eqn:potential_full}
\Phi_p (r, \theta, \phi)= -{{GM_p} \over {(|{\bf{r}}-{\bf{r_p}}|^2 + s^2)^{1/2}}},
\en
where $M_p$ is the mass of the planet, ${\bf r}$ and ${\bf r_p}$ are three-dimensional radius vectors of the center of the grid cell in question and of the planet, and $s$ is the smoothing length. Since the smoothing length in three-dimensional calculations is used only to avoid the  singularity in the potential on the grid scale, we adopt the cell diagonal size at the position of the planet for the smoothing length.  We insert the planet at $R_p = 90$~au with a fixed, circular orbit. We use three planet masses: $M_p = 0.5, 1$, and $2~M_{\rm Jup}$. We run simulations for 500 planetary orbits. The planet mass is linearly increased over the first 5 planetary orbits.

For the thermal evolution, we adopt an adiabatic equation of state with an adiabatic index $\gamma=1.4$. The gas pressure and the internal energy are related as $P=(\gamma-1)e$. In addition to the thermal energy evolution via $P{\rm d}V$ work, which is accounted for by the first term of the right-hand-side of Equation (\ref{eqn:energy}), cooling/heating of the gas is realized through the relaxation of the temperature $T_{\rm g}$ towards the initial temperature $T_{\rm g, init}$ (described in Equations \ref{eqn:temperature} - \ref{eqn:temperature_mid}) over the thermal relaxation timescale $t_{\rm relax}$ computed in Section \ref{sec:thermal_relaxation}. The thermal relaxation rate can be written as 
\be
\label{eqn:cooling_rate}
Q_{\rm relax} = - \rho_{\rm g} c_V  {{T_{\rm g} - T_{\rm g, init}} \over {t_{\rm relax}}}.
\en
In practice, we use $\max(t_{\rm relax}, \Delta t_{\rm hydro})$ in the denominator of Equation (\ref{eqn:cooling_rate}) where $\Delta t_{\rm hydro}$ is the hydrodynamic timestep, in order to avoid over-relaxation that can happen when $t_{\rm relax} < \Delta t_{\rm hydro}$.

For our standard model, we  take into account radiation, diffusion, gas-dust collision, and line cooling (Section \ref{sec:hydro_standard}). In practice, this is done by adopting a prescribed relaxation timescale using Equation (\ref{eqn:t_relax}), along with the fits in Equations (\ref{eqn:cooling_diff}) and (\ref{eqn:cooling_coll}). As a comparison to the standard model, we additionally carry out simulations without gas-dust collision: i.e., $t_{\rm relax} = {\rm max}(t_{\rm rad}, t_{\rm diff})\exp[-(Z/Z_{\rm line})^{-12}]$ (Section \ref{sec:hydro_te}; hereafter gas-dust thermal equilibrium model). This model is similar to what is often used in three-dimensional protoplanetary disk simulations where it is implicitly assumed that the gas and dust have instantaneous energy balance.

\subsection{Simulation Setup}
\label{sec:methods}

The simulation domain extends from 30~au (=$0.33~R_p$) to 210~au (=$2.33~R_p$) in $r$, from $\pi/2-0.4$ to $\pi/2$ in $\theta$ which covers 5.6 scale heights at the radial location of the planet, and from 0 to $2\pi$ in $\phi$. We adopt 460 logarithmically-spaced grid cells in the radial direction, 96 uniformly-spaced grid cells in the meridional direction, and 1482 uniformly-spaced grid cells in the azimuthal direction. With this choice, one gas scale height at the location of the planet is resolved with about 18 grid cells in all direction.

At the radial boundaries, we adopt a wave-damping zone to suppress wave reflection \citep{devalborro06}. At the lower meridional boundary, which is the disk midplane, we adopt the symmetric boundary condition for all variables but the meridional velocity for which we apply the reflecting boundary condition. At the upper meridional boundary we adopt the zero-gradient boundary condition. 

We adopt a kinematic viscosity characterized by $\alpha=10^{-3}$, a value broadly consistent with the level of turbulence observationally constrained for the TW~Hya disk \citep{teague16,flaherty18}. 

In order to ensure that no other hydrodynamic instabilities operate in the disk, we ran both standard model and gas-dust thermal equilibrium model in the absence a planet. We found that no hydrodynamic instabilities develop in these runs. The vertical shear instability \citep{urpin98,nelson13} is suppressed due to the non-zero viscosity and a long thermal relaxation time, in agreement with previous studies \citep{nelson13}.

\subsection{Simulation Results}
\label{sec:results}

\subsubsection{Standard Model}
\label{sec:hydro_standard}

We start by discussing the results from our standard model where we consider radiation, diffusion, gas-dust collision, and line cooling for the thermal relaxation of the disk gas.  Figure \ref{fig:hydro_wtcoll} shows the perturbed density, perturbed temperature, and vertical velocity in a $\phi-r$ planet at $Z/R = 0.23$ for the $M_p = 0.5~M_{\rm Jup}$ model. This height corresponds to about three scale heights above the midplane at the radial location of the planet and is close to the $^{12}$CO emission surface in the synthetic observation we will present in Section \ref{sec:simobs}. 

As most clearly shown in the perturbed density distribution, the planet excites a pair of primary Lindblad spirals, one in the inner disk (the arc crossing the $r=40$~au boundary at $\phi \sim 3\pi/4$) and one in the outer disk (the arc crossing the $r=140$~au boundary at $\phi \sim -\pi/4$).  The Lindblad spirals are nearly perpendicular to the azimuth axis near the planet, suggesting that they have a large pitch angle close to 90$^\circ$ there. In addition to the primary Lindblad spirals, the planet excites a secondary Lindblad spiral in the inner disk, which emerges at $r\sim50$~au and $\phi \sim \pi$. At $Z/R=0.23$, density, temperature, vertical velocity perturbations created by the Lindblad spirals of the $0.5~M_{\rm Jup}$ planet are a few to $10~\%$ of the background values or the local sound speed. 

Along with the Lindblad spirals, the planet excites a family of spirals via buoyancy resonances, which can be  most clearly seen in the vertical velocity plot. Two main features that distinguish buoyancy spirals from Lindblad spirals are (1) the tightly wound morphology and (2) the large vertical motions, which we will explain in detail one by one.
  
\begin{figure}[t]
    \centering
    \includegraphics[width=0.48\textwidth]{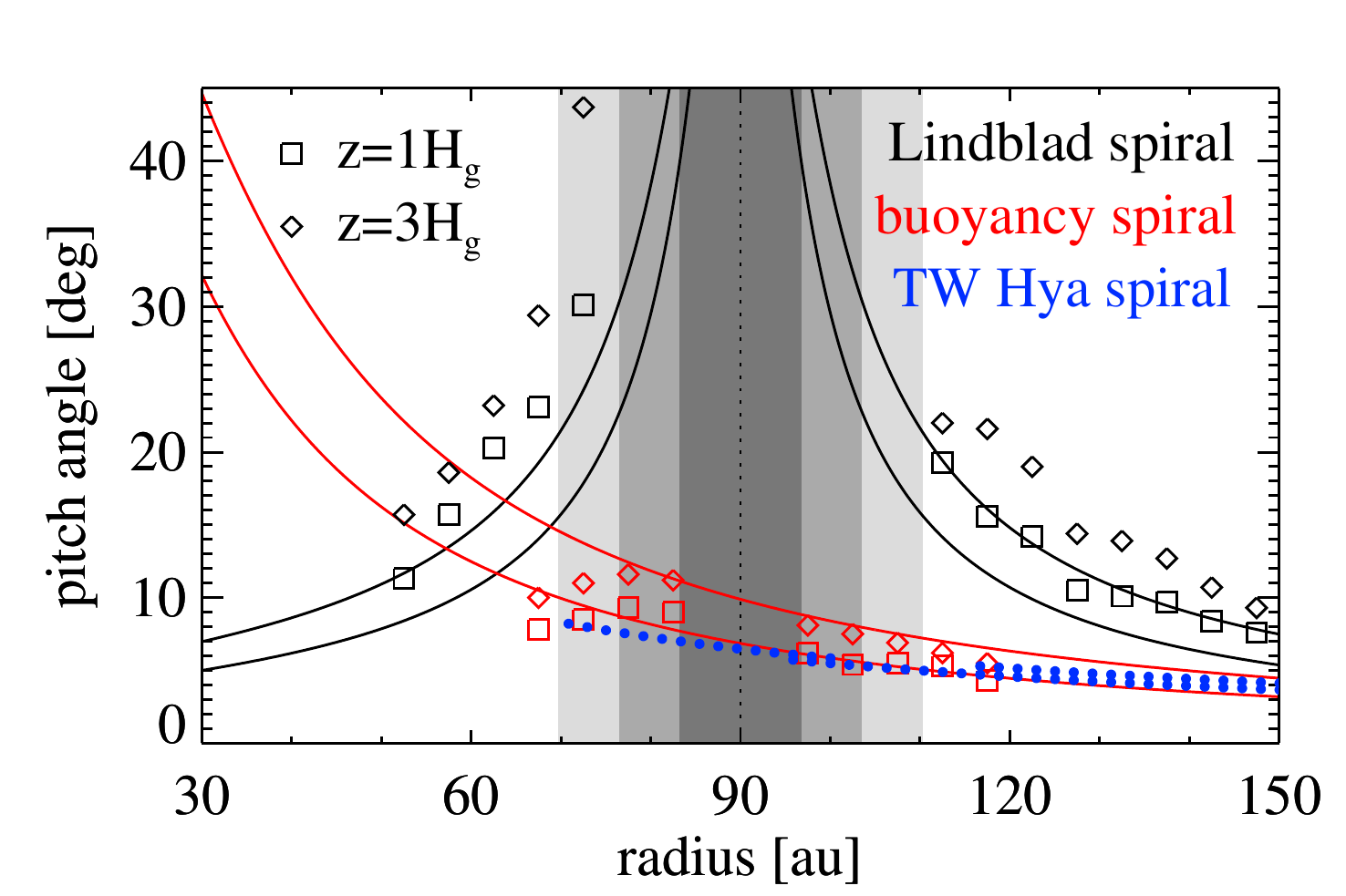}
    \caption{The pitch angles of (black curves) Lindblad and (red curves) buoyancy spirals calculated based on the linear theory. For each spirals, the lower and upper curves show the linear theory prediction at $Z=1~H_{\rm g}$ and $3~H_{\rm g}$, respectively. Square and diamond symbols present measured pitch angles from the hydrodynamic simulation with $M_p = 0.5~M_{\rm Jup}$ at (square) $Z=1~H_{\rm g}$ and (diamond) $3~H_{\rm g}$. Blue dots present the pitch angle of the spirals observed in the TW~Hya disk \citep{teague19}. The gray shaded regions show $\pm$1, $\pm$2, and $\pm$3 scale heights from the planet in the radial direction.}
    \label{fig:pitchangle}
\end{figure}

In Figure \ref{fig:pitchangle}, we show the measured pitch angle of Lindblad and buoyancy spirals with a 5~au interval in radius. For Lindblad spirals, we measure the pitch angle using the peak in the density perturbation. For buoyancy spirals, we measure the pitch angle using the peak in positive vertical velocities. We opt to use the vertical velocity instead of the density because (1) buoyancy spirals are more clearly identifiable with the vertical velocity and (2) we can make a more direct comparison to the TW~Hya spiral detected in the velocity space (Section \ref{sec:twhya}). 

We also present the pitch angle derived with linear theory. For Lindblad spirals, the phase angle $\phi_{\rm L}$ (i.e., the azimuthal angle from the spiral to the planet) as a function of the cylindrical disk radius $R$ is 
\begin{eqnarray}
\label{eqn:lindblad_phase}
\phi_{\rm L}(R) = &-& {\rm sgn}(R - R_p) { \pi \over 4m}
\nonumber\\
&-& \int_{R_m^\pm}^{R} {\Omega(R') \over c_s(R')} \left| \left(1- {R'^{3/2} \over {R_p^{3/2}}} \right)^2 - {1 \over m^2} \right|^{1/2} {\rm d}R'\phantom{1111}
\end{eqnarray}
following \citet{baezhu18a}, where $m = (1/2)(H/R)_p^{-1}$ and $R_{m}^{\pm}=({ 1\pm 1/m})^{2/3} R_p$.

For buoyancy spirals, the phase angle $\phi_{\rm B}$ is
\begin{eqnarray}
\label{eqn:buoyancy_phase}
\phi_{\rm B} (R) & = &\pm 2n\pi {{\left| \Omega_p-\Omega_K \right|} \over \Omega_K} {H_{\rm g} \over Z} \left(1+{Z^2 \over R^2} \right)^{3/2} \\
\nonumber
& & \times \left[  {\gamma -1 \over {\gamma}} + { 2 c_s \over g} {\partial c_s \over \partial Z} \right]^{-1/2},
\end{eqnarray}
following \citet{zhu15}, where $n$ is a positive integer and $g$ is the gravity from the star. Compared with the analytic form given in \citet{zhu15}, note that Equation (\ref{eqn:buoyancy_phase}) has an additional term in the parenthesis of the right-hand-side, ($2c_s/g$)$\partial c_s/\partial Z$, which takes the vertical temperature stratification into account. When there is no vertical temperature stratification Equation (\ref{eqn:buoyancy_phase}) reduces to the phase angle derived in \citet{zhu15}. The pitch angles of Lindblad and buoyancy spirals are computed as $\beta \equiv -\tan^{-1}({\rm d}R/R{\rm d}\phi)$, using the phase equations in Equations (\ref{eqn:lindblad_phase}) and (\ref{eqn:buoyancy_phase}).

\begin{figure*}[]
    \centering
    \includegraphics[width=0.95\textwidth]{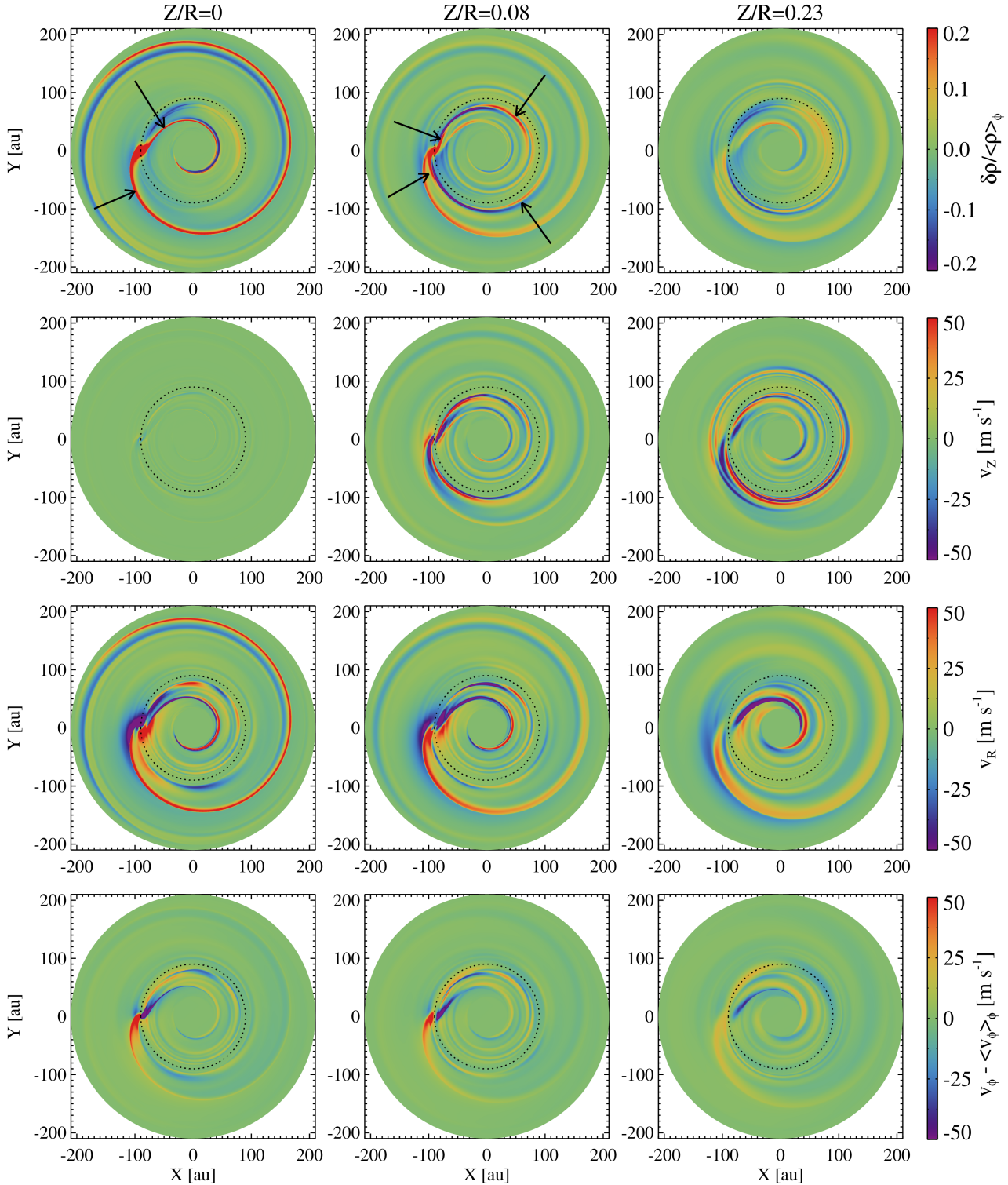}
    \caption{The two-dimensional distributions of (first row) perturbed density $\delta\rho/\langle \rho \rangle_\phi$, (second row) vertical velocity $v_Z$, (third row) radial velocity $v_R$, and (fourth row) perturbed azimuthal velocity $v_\phi - \langle v_\phi \rangle_\phi$, taken at the end of the simulation with $M_p = 0.5~M_{\rm Jup}$. From left to right in each row, three panels present the distributions at $Z/R=0$ (the midplane), at $Z/R=0.08$ ($\simeq 1 H$ at 90~au), and $Z/R=0.23$ ($\simeq 3 H$ at 90~au), respectively. The two arrows in the upper left panel point to the inner and outer Lindblad spirals, while the four arrows in the upper middle panel point to the first- and second-order, inner and outer buoyancy spirals. Dotted circles show the planet's orbit. The buoyancy spirals can be best distinguished from Lindblad spirals using vertical velocities.}
    \label{fig:vcomp}
\end{figure*}
\begin{figure*}[ht!]
    \centering
    \includegraphics[width=1\textwidth]{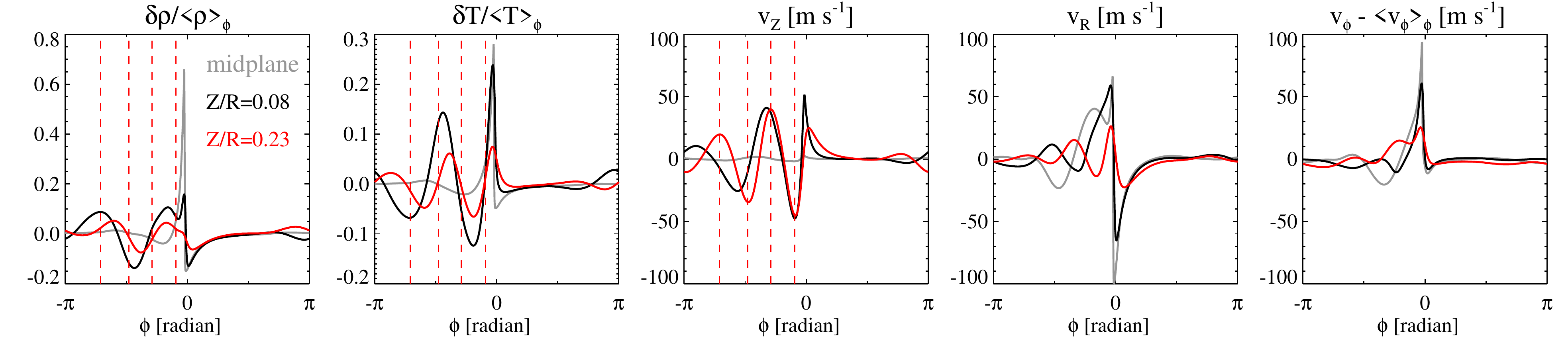}
    \includegraphics[width=1\textwidth]{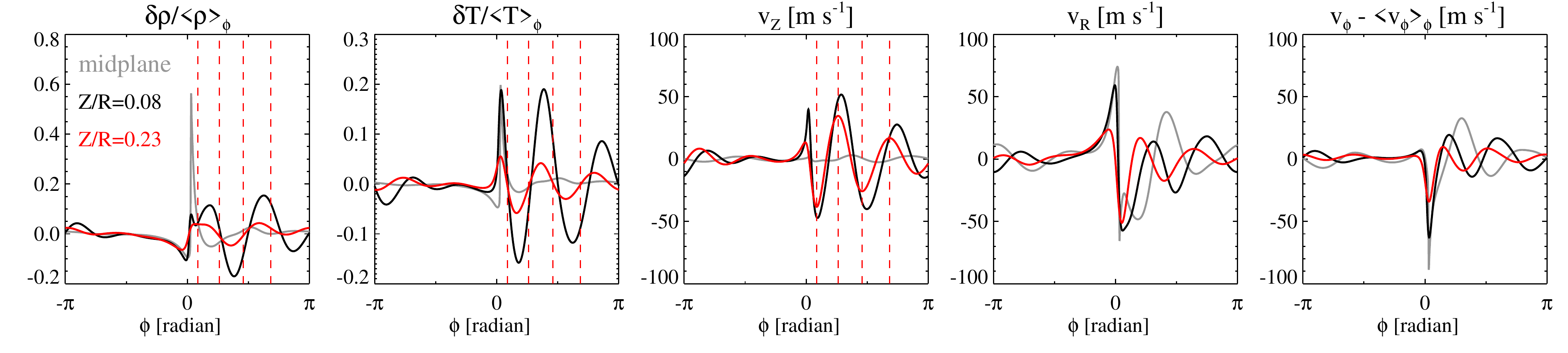}
    \caption{The azimuthal profiles of (from left to right) perturbed density $\delta\rho/\langle \rho \rangle_\phi$, perturbed temperature $\delta T / \langle T \rangle_\phi$, vertical velocity $v_Z$, radial velocity $v_R$, and perturbed azimuthal velocity $v_\phi - \langle v_\phi \rangle_\phi$ from the standard model with $M_p=0.5~M_{\rm Jup}$. The upper panels present the profiles at two scale heights beyond the planet ($R=$103.5~au) and the lower panels present the profiles at two scale heights inward of the planet ($R=$76.5~au). In each panel, gray, black, and red curves present the azimuthal profiles at the midplane, at $Z/R=0.08$ ($\simeq 1 H_{\rm g}$ at 90~au), and $Z/R=0.23$ ($\simeq 3 H_{\rm g}$ at 90~au), respectively. The red dashed lines show the azimuthal locations where the vertical velocity peaks.}
    \label{fig:azimuth_wtcoll}
\end{figure*}

As shown in Figure \ref{fig:pitchangle}, buoyancy spirals' pitch angle is smaller than Lindblad spirals' pitch angle over a broad range of distance from the planet. In particular, within a few scale heights from the planet Lindblad spirals' pitch angle increases to $90^\circ$ toward the planet, whereas buoyancy spirals' pitch angle remains $\lesssim 10^\circ$. We note that Figure \ref{fig:pitchangle} shows pitch angles of the first-order buoyancy spirals only (i.e., $n=1$ in Equation \ref{eqn:buoyancy_phase}) for both linear theory prediction and measurement from the simulations. For higher-order buoyancy spirals (i.e., $n > 1$), the pitch angle is smaller than that of first-order buoyancy spirals by a factor of $\simeq 1/n$, meaning that they are more tightly wound.

It is also worth pointing out that buoyancy spirals' pitch angle varies continuously as a function of radius without a singularity at the radial location of the planet, unlike Lindblad spirals. From the observational point of view, this implies that the inner and outer buoyancy spirals can appear connected to each other as a single spiral, especially when the spatial resolution is insufficient.

Another characteristic that distinguishes buoyancy spirals from Lindblad spirals is the large vertical motions. In order to visualize the density and velocity perturbations at different heights, we present the perturbed density and vertical, radial, azimuthal velocities at the midplane, $Z/R=0.08$, and $Z/R=0.23$ in Figure \ref{fig:vcomp}. The azimuthal profiles of the perturbed density, perturbed temperature, and vertical, radial, and azimuthal velocities at $R= R_p\pm2~H_{\rm g}$  are presented in Figure \ref{fig:azimuth_wtcoll}. 

Focusing on the Lindblad spirals first, both density and velocity perturbations driven by Lindblad spirals is the strongest at the midplane and decreases over height. 
The only exception is the near-zero vertical velocity at the midplane, which is because of the symmetry across the midplane. 
For the buoyancy spirals, on the other hand, perturbations are the smallest at the midplane because $N_Z = 0$ there. Between $Z/R=0.08$ and 0.23, perturbations remain comparable or become stronger than their Lindblad counterparts. In particular, we note that the vertical velocity perturbation associated with buoyancy spirals become stronger and more extended in azimuth over height. Compared to Lindblad spirals, buoyancy spirals generally produce smaller perturbations, but the vertical motions associated with buoyancy spirals can be stronger especially as we move to the surface layers. These characteristics of buoyancy spirals suggest that the best strategy to observe buoyancy spirals is to look for vertical velocity perturbations in the surface layers of face-on disks.

Finally, we note that there is a $\pi/2$ phase shift between the vertical velocity and density/temperature perturbations driven by buoyancy resonances. At $R=103.5$~au, for example, the vertical velocity has peaks at $\phi = -0.29, -0.91, -1.50$, and $-2.23$~radians and the density and temperature perturbation is close to zero at those azimuthal locations (see the red dashed lines in Figure \ref{fig:azimuth_wtcoll}).

\subsubsection{Gas-dust Thermal Equilibrium Model}
\label{sec:hydro_te}

We now turn our discussion to the gas-dust thermal equilibrium model in which the gas is assumed to be thermally coupled with the dust. With this assumption the relaxation timescale in the surface layers is much shorter than the dynamical time (Figure \ref{fig:cooling_time}c) and buoyancy resonances are expected to be weak or absent in the surface layers (Figure \ref{fig:t_relax_Nz} right).

\begin{figure*}[ht!]
    \centering
    \includegraphics[width=1\textwidth]{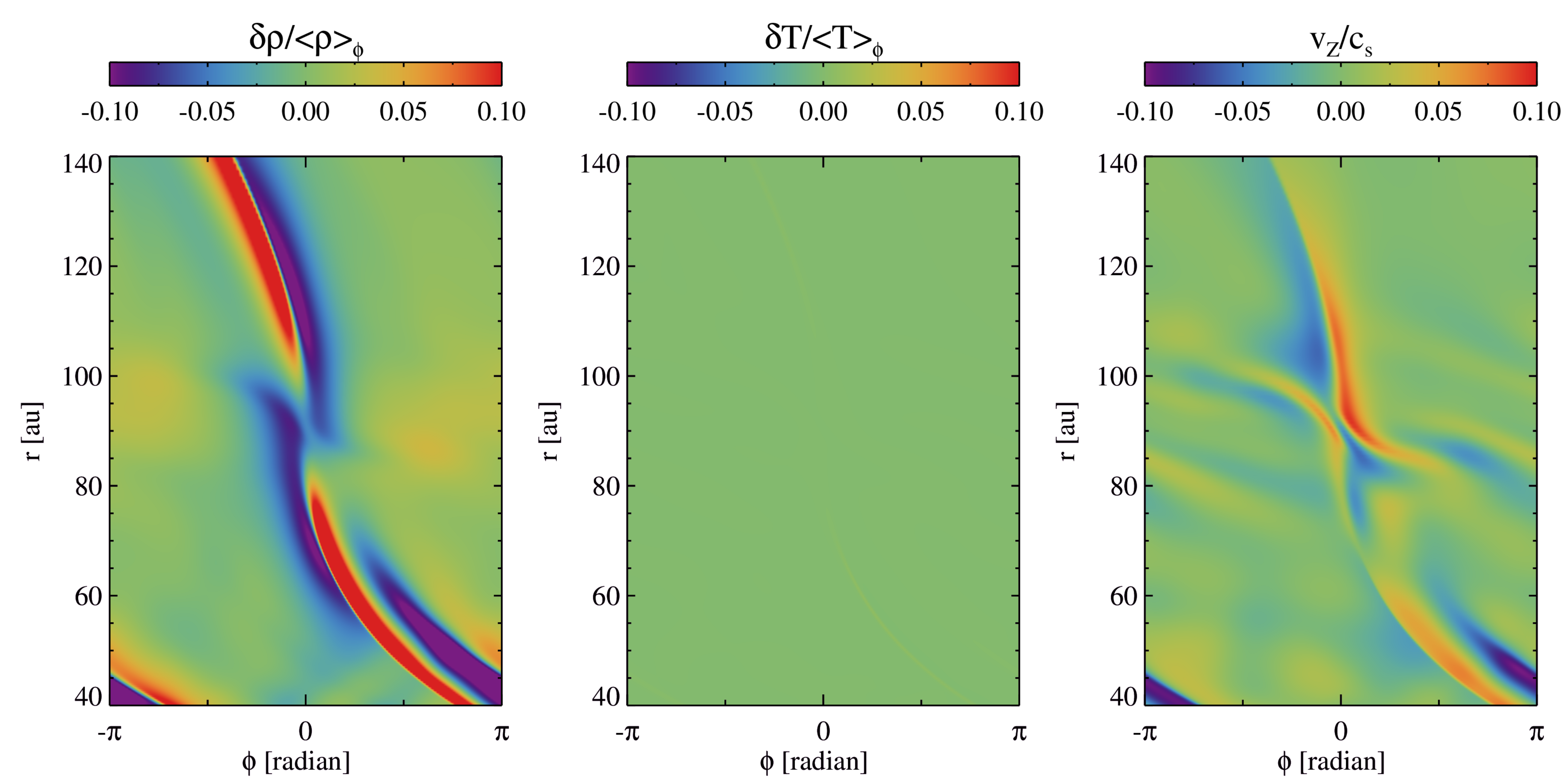}
    \caption{
        Same as Figure \ref{fig:hydro_wtcoll}, but for the gas-dust thermal equilibrium model. Due to the short thermal relaxation timescale in the surface layers, buoyancy resonances do not fully develop.}
    \label{fig:hydro_wotcoll}
\end{figure*}
\begin{figure*}[]
    \centering
    \includegraphics[width=1\textwidth]{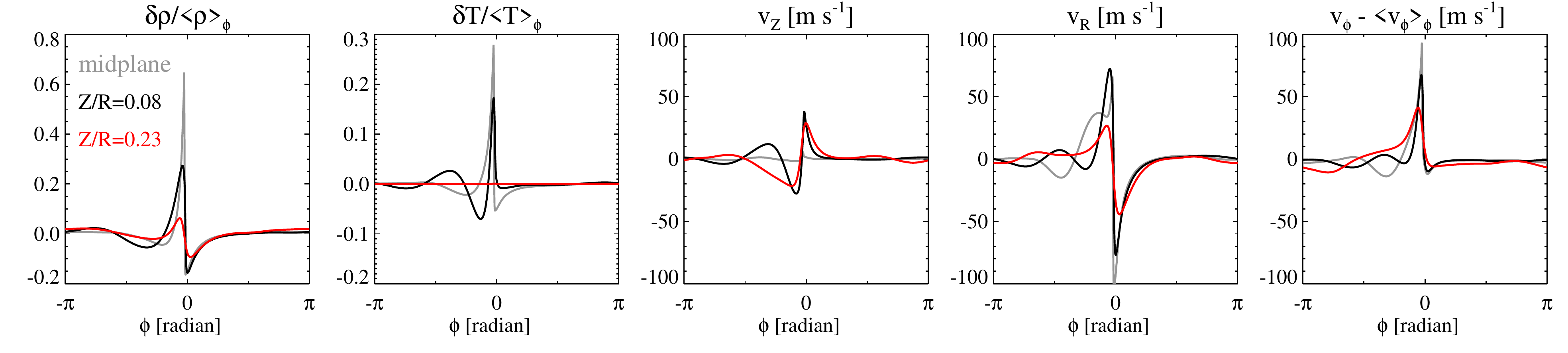}
    \includegraphics[width=1\textwidth]{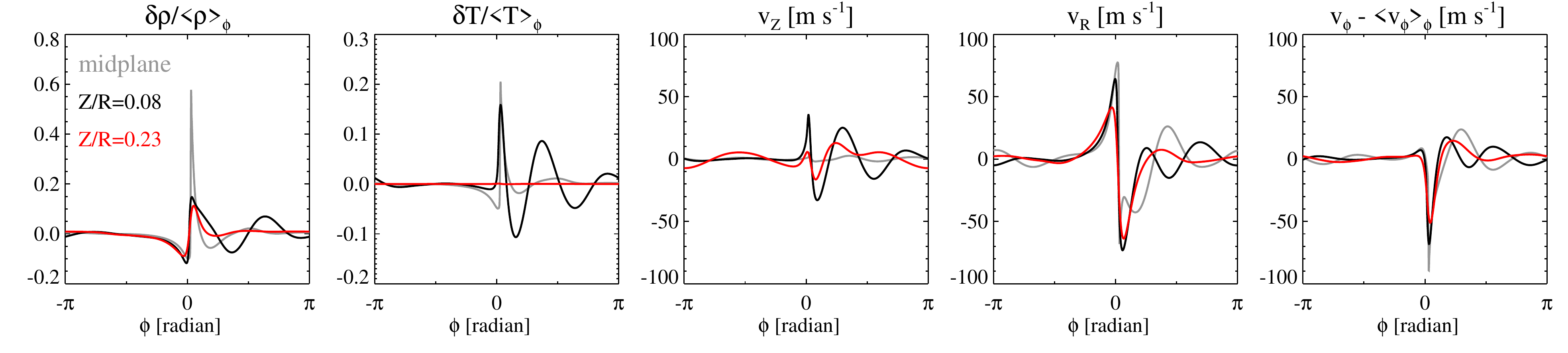}
    \caption{Same as Figure \ref{fig:azimuth_wtcoll}, but for the gas-dust thermal equilibrium model. 
    }
    \label{fig:azimuth_wotcoll}
\end{figure*}

The perturbed density, perturbed temperature, and vertical velocity at $Z/R=0.23$ is presented in Figure \ref{fig:hydro_wotcoll}. In Figure \ref{fig:azimuth_wotcoll} we present the azimuthal profiles of the perturbed density, perturbed temperature, and vertical, radial, and azimuthal velocities at $R = R_p\pm2H_{\rm g}$ from the planet in the radial direction. As apparent from the figures, buoyancy spirals do not generate clear density and temperature perturbations. Vertical velocity perturbations arising from buoyancy resonances are seen, but they are much weaker compared with the standard model. We note that the resonance is not completely suppressed in this model, as suggested by the vertical velocity perturbation, because the buoyancy frequency is not strictly zero with the imposed stratified disk temperature (see Equation \ref{eqn:buoyancy_frequency_ideal}).

Speaking of Lindblad spirals, we find that they produce smaller perturbations as we move toward the surface layers, in agreement with what is seen in the standard model. However, we note that the level of density perturbations at $Z/R=0.08$ and $0.23$ are larger than the standard model by about a factor of two. Note also that the Lindblad spirals are more tightly wound -- compare the azimuthal angles where Lindblad spirals meet $r=40$ and 140~au in Figure \ref{fig:hydro_wtcoll} and \ref{fig:hydro_wotcoll}. This is because the gas behaves (nearly) isothermally and thus the sound speed is smaller than the standard model where the gas behaves adiabatically.

\section{Simulated Observations}
\label{sec:simobs}

In order to examine the observability of buoyancy spirals, we carry out synthetic observations of the $^{12}$CO $J = 3-2$ and $^{13}$CO $J = 3-2$ lines using the 3D radiative transfer code RADMC-3D \citep{radmc3d}. To do so, we first add an inner disk inward of the computational domain, from 1 to 30~au, using the gas density profile described with Equation (\ref{eqn:sigma}). We assume CO is photodissociated at the surface layers, where the sum of the vertically\footnote{In practice, this is done along the meridional direction in the spherical coordinates.} integrated gas column density (to consider external irradiation) and the radially integrated gas column density from the central star (to consider the central star's irradiation) is less than $10^{21}~{\rm cm}^{-2}$ \citep{visser09}. We assume CO is frozen onto grains in the regions where the temperature is below 21~K \citep{schwarz16}. We adopt a $^{12}$CO-to-H$_2$ ratio of $10^{-6}$, which is smaller than the canonical interstellar medium value of $10^{-4}$, motivated by the fact that CO in the TW~Hya is known to be largely depleted in the gas phase \citep{schwarz16,zhang19}. For $^{13}$CO, we adopt $^{13}$CO/$^{12}$CO ratio of 1/60.

As opposed to running Monte Carlo calculations to compute the dust temperature and assuming the gas and dust are thermally coupled, an approach often taken in the post-processing of hydrodynamic simulations, we adopt the gas temperature from the hydrodynamic simulations. This is because the gas and dust do not necessarily have the same temperature as we discussed in Section \ref{sec:thermodynamics}. 

\begin{figure*}[]
    \centering
    \includegraphics[width=1.05\textwidth]{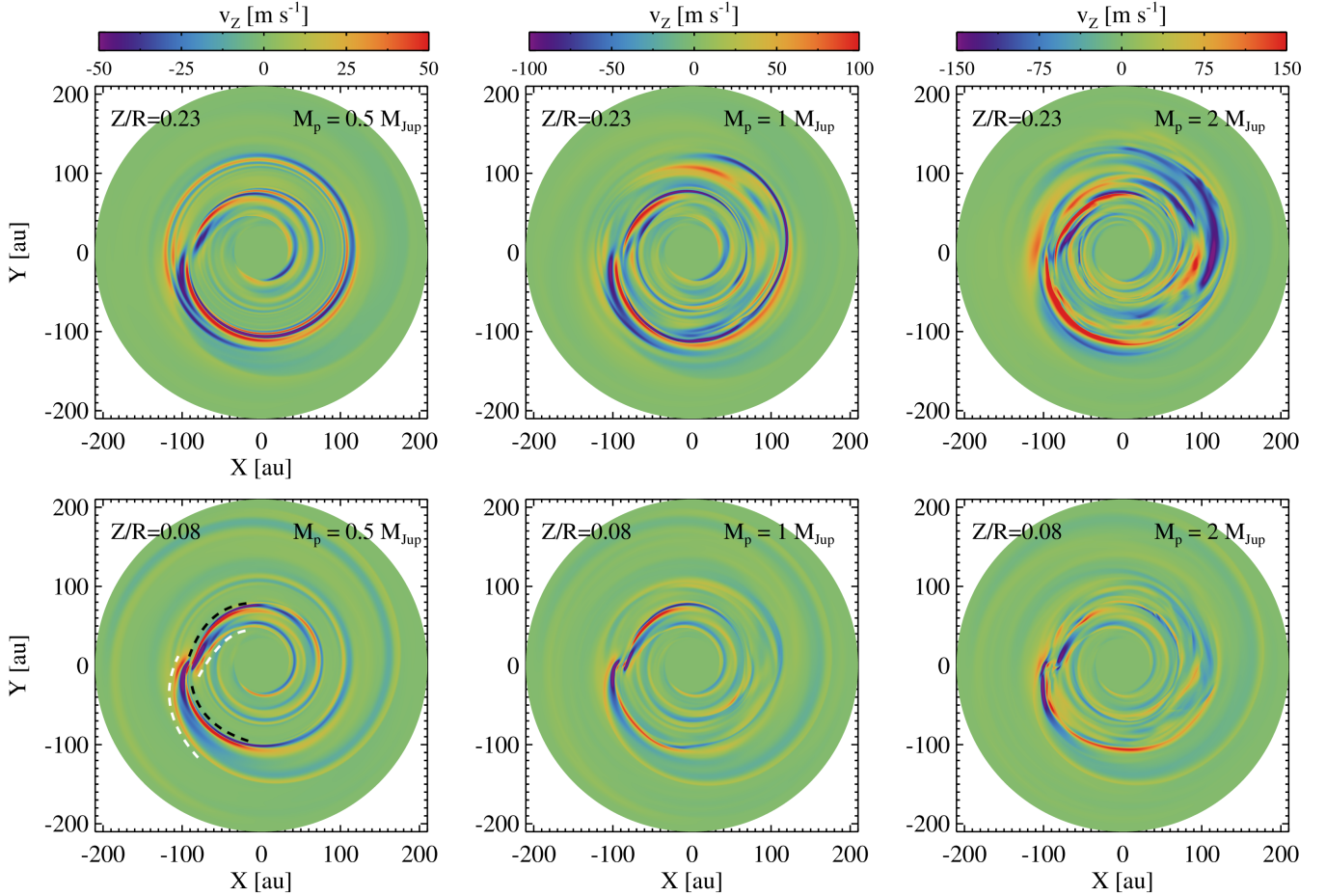}
    \caption{The vertical velocity $v_Z$ in a $X-Y$ plane in units of ${\rm m~s}^{-1}$ at (upper panels) $Z/R=0.23$ ($\simeq 3~H_{\rm g}$ at 90~au) and (lower panels) $Z/R=0.08$ ($\simeq 1~H_{\rm g}$ at 90~au), after 500 orbits. From left to right, results from standard 0.5~$M_{\rm Jup}$, 1~$M_{\rm Jup}$, and 2~$M_{\rm Jup}$ models, respectively. The planet is located at $(X, Y)=(-90~{\rm au}, 0~{\rm au})$. Note that the vertical motions driven by Lindblad spirals are smaller than that those driven by buoyancy spirals, especially at $Z/R=0.23$. The black and white dashed curves in the lower left panel trace the buoyancy and Lindblad spirals, respectively.}
    \label{fig:hydroxy}
\end{figure*}
\begin{figure*}[]
    \centering
    \includegraphics[width=\textwidth]{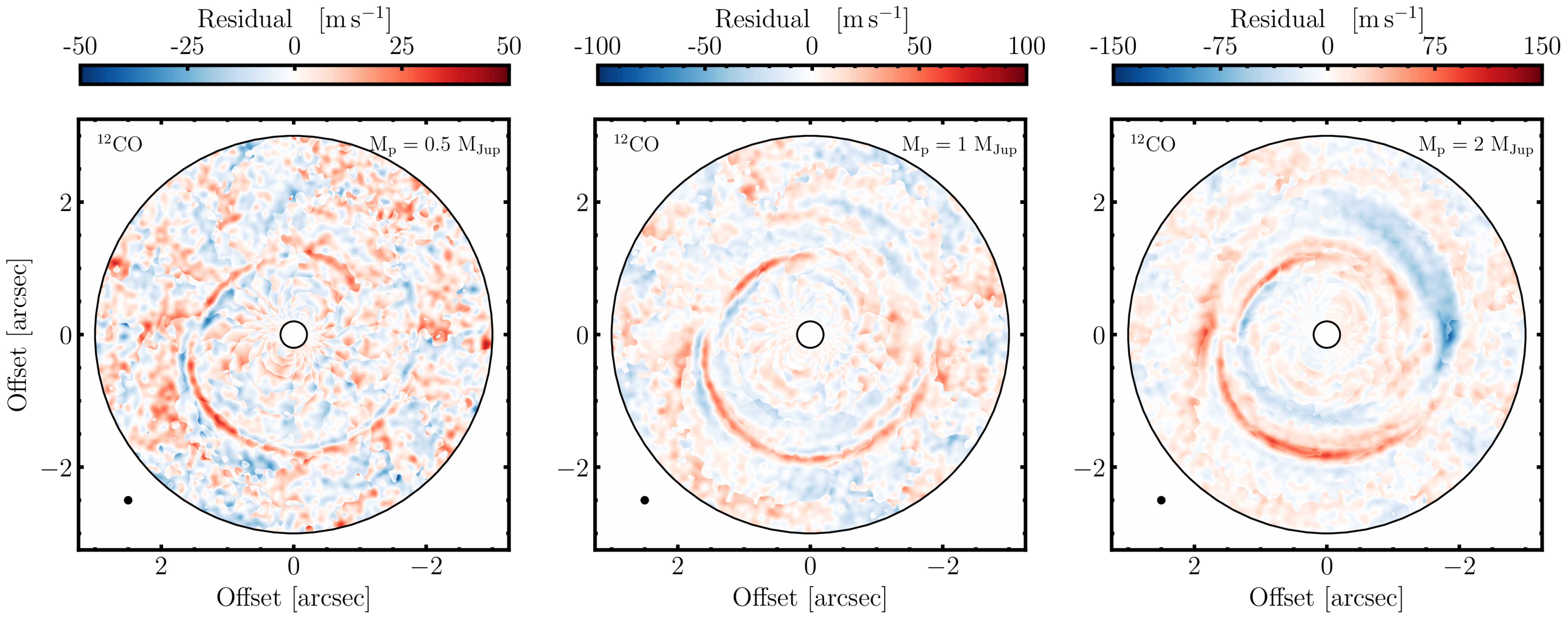}
        \includegraphics[width=\textwidth]{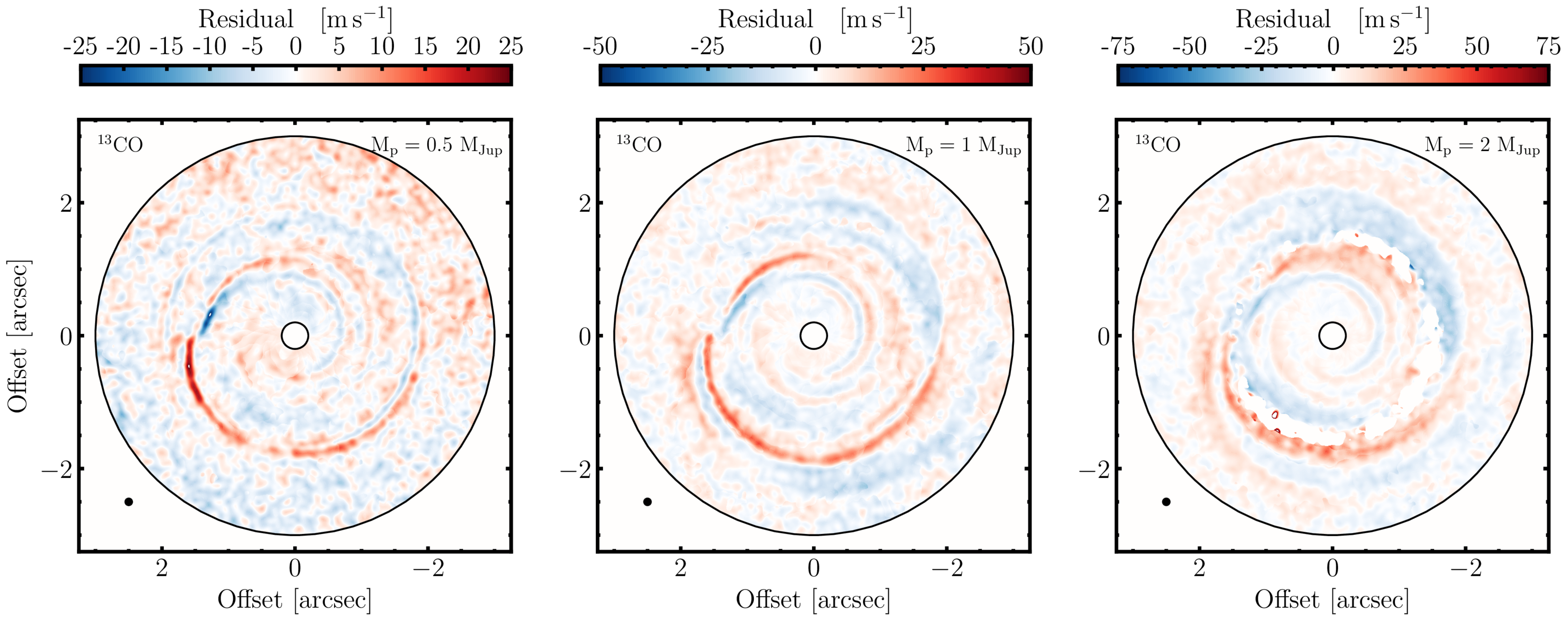}
    \caption{Keplerian-subtracted moment maps with simulated (upper panels) $^{12}$CO and (lower panels) $^{13}$CO observations adopting 0.1'' spatial resolution, 100~${\rm m~s}^{-1}$ spectral resolution, and 1.0~K and 0.3~K rms noise level, respectively. From left to right, results with standard 0.5~$M_{\rm Jup}$, 1~$M_{\rm Jup}$, and 2~$M_{\rm Jup}$ models. The beam is shown in the lower left corner of each panel.}
    \label{fig:simobs}
\end{figure*}

For the fiducial disk geometry, we use a disk position angle (PA) of $152^\circ$, defined as the angle from the north to the redshifted major axis in the counter-clockwise direction, and an inclination of $5^\circ$, comparable to those of the TW~Hya disk. 
We test the influence of varying disk inclination on the kinematic signatures of buoyancy spirals in Section \ref{sec:inclination}. The planet is placed at PA=$90^\circ$ (i.e., East) in all cases. The disk rotates clockwise on the sky.

We create image cubes at 10~${\rm m~s}^{-1}$ velocity resolution and average down to the desired velocity resolution of 100~${\rm m\,s}^{-1}$. This is because most radiative transfer codes including RADMC-3D return emission at the central frequency of each channel, rather than the integrated emission across the channel (see \citealt{rosenfeld13} for a demonstration of this effect). We then convolve the image cubes with a circular Gaussian beam with a FWHM of $0.1\arcsec$. Observations at this high spatial resolution will be the product of multiple executions with differing array configurations and observing conditions in order to fill in the $uv$ plane. As we are not trying to recreate specific observations, where this would be a necessary step, but rather provide a quantitative prediction, we opt to simply convolve each channel with a circular Gaussian beam. We add correlated (both spatially due to the beam and spectrally due to the Hanning smoothing) noise to each channel with specified RMS of 1~K for $^{12}$CO and 0.3~K for $^{13}$CO, which correspond to 0.94 and 0.28~mJy~beam$^{-1}$, respectively.

We provide simulated $^{12}$CO and $^{13}$CO cubes (averaged down to 100 ${\rm m~s}^{-1}$ velocity resolution but without beam convolution and correlated noise) at \url{https://doi.org/10.5281/zenodo.4361639}.

\subsection{Buoyancy Spirals in Keplerian-subtracted Moment Maps}

Before we present Keplerian-subtracted moment maps, we show the vertical velocity distribution from standard $0.5~M_{\rm Jup}$, $1~M_{\rm Jup}$, and $2~M_{\rm Jup}$ models in Figure \ref{fig:hydroxy}. Buoyancy resonances produce vertical motions of order of 100~m~s$^{-1}$, with a larger magnitude for more massive planets. As discussed earlier, Lindblad spirals produce much smaller vertical perturbations compared with buoyancy spirals, especially in the surface layers. We thus do not expect to detect them in $^{12}$CO in face-on disks.

We generate Keplerian-subtracted moment maps using the quadratic method described in \citet{Teague_Foreman-Mackey_2018}. Using the Python package \texttt{eddy} \citep{eddy}, we fit a Keplerian rotation profile to the rotation maps, allowing the source center, the disk inclination and position angle and the systemic velocity to vary. Given the low inclination of TW~Hya ($i \sim 5\degr$) we do not include any terms describing an elevated emission surface. 

\begin{figure*}[ht!]
    \centering
    \includegraphics[width=\textwidth]{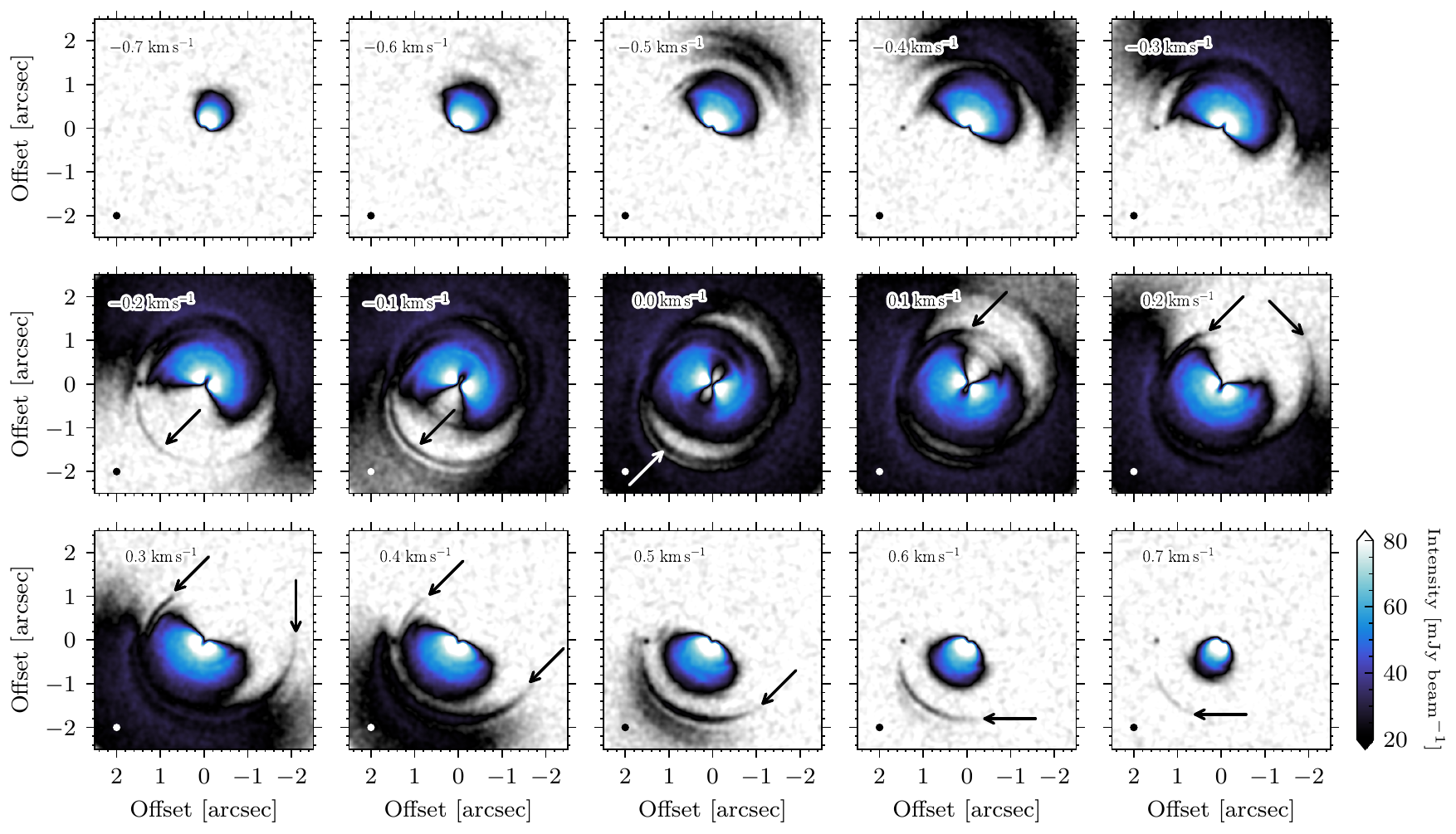}
    \caption{Channel maps from a simulated $^{12}$CO (3-2) line observation based on the hydrodynamic simulation with 2~$M_{\rm Jup}$ planet. The channel velocity, relative to the central channel, is presented in the upper left corner of each panel. Buoyancy spirals are pointed out with white/black arrows in relevant panels. The beam is presented in the lower left corner of each panel.}
    \label{fig:channelmap}
\end{figure*}

The Keplerian-subtracted moment maps are shown in Figure~\ref{fig:simobs}. We note that the residual maps clearly show a coherent, tightly wound spiral structure. It is also worth pointing out the excellent agreement between the input velocity field from hydrodynamic simulations and the retrieved velocities.

\begin{figure*}[]
    \centering
    \includegraphics[width=\textwidth]{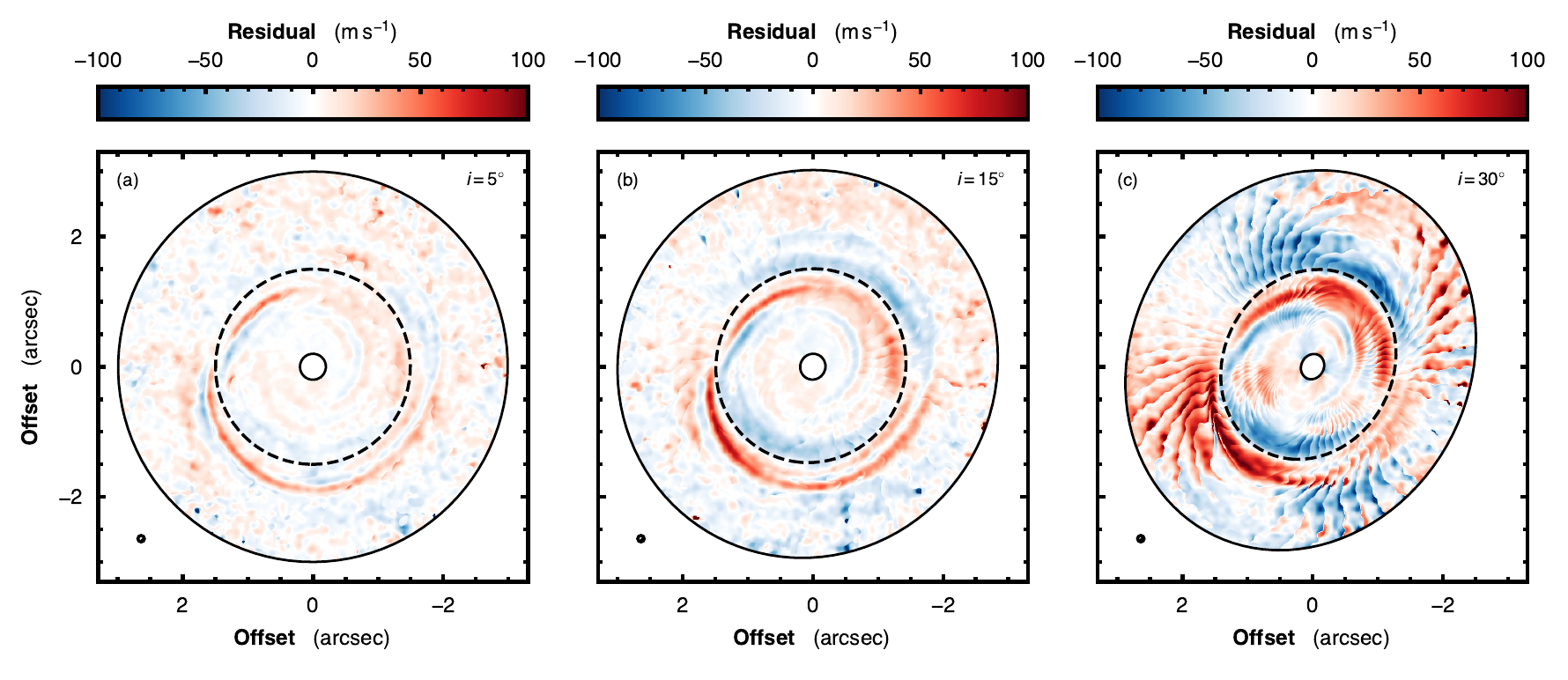}
    \caption{Keplerian-subtracted moment maps with (a) 5, (b) 15, and (c) 30 degrees inclination for the $1~M_{\rm Jup}$ model. The dotted ellipse in each panel shows planet's radial location $r=1.5''$ on sky. The feather-like feature in  panel (c) is due to the channelization effect owing to the channel spacing of the data. Higher spectral resolution data can suppress this effect (see \citet{christiaens14} for an example).}
    \label{fig:simobs_inc}
\end{figure*}
\begin{figure*}[ht!]
    \centering
    \includegraphics[width=\textwidth]{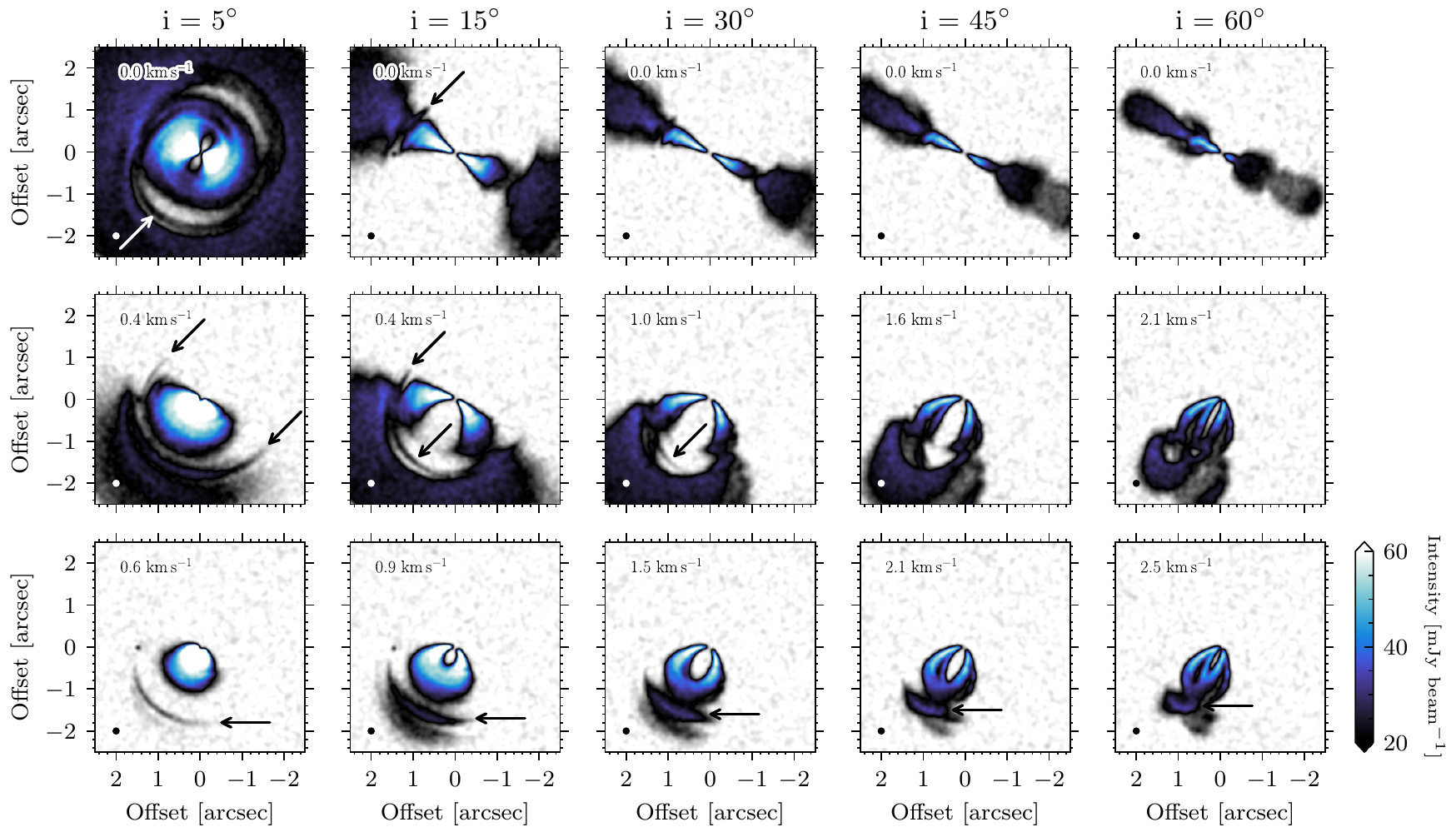}
    \caption{Channel maps from a simulated $^{12}$CO (3-2) line observation based on the hydrodynamic simulation with 2~$M_{\rm Jup}$ planet, using various disk inclination: (from left to right) 5, 15, 30, 45, and 60 degrees. The channel velocity, relative to the central channel, is presented in the upper left corner of each panel in units of ${\rm km~s}^{-1}$. Buoyancy spirals are pointed out with white/black arrows in relevant panels. The  beam is shown in the lower left corner of each panel.}
    \label{fig:channel3_2mjup}
\end{figure*}

\subsubsection{Case Study: TW Hya}
\label{sec:twhya}

Interestingly enough, the tightly wound morphology of buoyancy spirals resemble to the spiral seen in the Keplerian-subtracted moment map of TW~Hya  \citep{teague19}. As presented in Figure \ref{fig:pitchangle} the buoyancy spirals driven by a planet at 90~au can explain the small pitch angle as well as the monotonically decreasing pattern over radius. The magnitude of observed velocity perturbations ($\sim40~{\rm m~s}^{-1}$) is also in a good agreement with our simulations.

We note that it is not impossible to explain the small pitch angle of the TW~Hya spirals with Lindblad resonance; one may place a planet far inward of the observed spirals. However, if this has to be the case, it is unclear why we do not see any large perturbations in the inner disk from neither line nor continuum observations because we expect the largest perturbations would arise at the vicinity of the planet (Figure \ref{fig:vcomp}). In addition, given the low inclination of the TW~Hya disk we are likely seeing vertical motions. Our simulations show that the outer Lindblad spiral produces little vertical motions ($\ll 50~{\rm m~s}^{-1}$) and it is unlikely that the velocity spiral in the TW~Hya disk is associated with Lindblad spirals. 

We thus propose that, if the spirals in TW~Hya are driven by an embedded planet, the spirals could be associated with buoyancy resonances driven by a (sub-)Jovian-mass planet at around 90~au.

\subsection{Buoyancy Spirals in Velocity Channel Maps}

When the perturbations driven by buoyancy spirals are sufficiently large, the spirals can be seen in velocity channel maps. We present channel maps of the synthetic $^{12}$CO line observation from the standard $2~M_{\rm Jup}$ model in Figure \ref{fig:channelmap}. Channel maps from standard $0.5~M_{\rm Jup}$ and $1~M_{\rm Jup}$ models are presented in Figures \ref{fig:channel3_05mjup} and \ref{fig:channel3_1mjup} in Appendix \ref{sec:channel_maps}. 

The main characteristic of buoyancy spirals in channel maps is wedge-like features standing out of the so-called butterfly pattern of the Keplerian disk. In Figure \ref{fig:channelmap} these non-Keplerian features are most clearly seen close to the planet, on the North-East side of the disk at $v=0.2 - 0.3~{\rm km~s}^{-1}$. Along the major-axis of the disk, buoyancy spirals can appear as an arc bridging the Keplerian wings (e.g., South-East side at $v= -0.1~{\rm km~s}^{-1}$, North-West side at $v=0.1~{\rm km~s}^{-1}$) or as an arc beyond the inner disk (e.g., South-East side at $v= 0.5-0.6~{\rm km~s}^{-1}$).

\subsection{Effect of Disk Inclination}
\label{sec:inclination}

In order to examine the observational appearance of buoyancy spirals in more inclined disks, we generate additional image cubes adopting 15, 30, 45, and 60~degrees inclinations. Keplerian-subtracted moment maps for standard $1~M_{\rm Jup}$ model are shown in Figure \ref{fig:simobs_inc}. 

Although non-Keplerian velocity components arising from buoyancy resonances are still present, they appear more as an ellipse for more inclined geometry, making it challenging to identify buoyancy spirals. For $i=15^\circ$ and $i=30^\circ$ cases, the red-shifted buoyancy spirals appear stronger than $i=5^\circ$ case. This is because we are more sensitive to in-plane velocity perturbations for inclined disks. As the planet opens a gap around its orbit it alters the gas pressure profile from the unperturbed one, such that the disk gas rotates at sub-Keplerian speed inside of the planet's orbit and at super-Keplerian outside of the planet's orbit \citep{teague18}. For an inclined disk, sub-Keplerian rotation would appear as a blue-/red-shifted semi-ellipse on the red-/blue-shifted side of the disk, while super-Keplerian rotation would appear as a red-/blue-shifted semi-ellipse on the red-/blue-shifted side of the disk (see Figure 5 of \citealt{teague19}). The pattern we see across the planet's orbit in the Keplerian-subtracted moment maps in Figure \ref{fig:simobs_inc} exactly matches with the aforementioned expectation. If we first look at the $i=15^\circ$ model, there is an elongated blue-shifted arc on the red-shifted side (South-East side) of the disk, right inward of the planet's orbit. On the blue-shifted side of the disk, we see an elongated red-shifted arc inward of the planet's orbit, along with the red-shifted buoyancy spiral. Right outside of the planet's orbit on the red-shifted side, the super-Keplerian rotation adds red-shifted residuals to the buoyancy spiral, enhancing the overall magnitude of the residual. The same pattern is consistently observed in the $i=30^\circ$ model and the rotation modulation is stronger in this case.

It is thus reasonable to conclude that disentangling the vertical motions induced by buoyancy resonances and the modulation in rotational motions associated with the gap is generally more challenging for inclined disks. However, if a buoyancy spiral is sufficiently extended in azimuth such that the spiral crosses the minor axis of the disk, this offers a possibility to disentangle vertical motions from rotational motions. This is because vertical motions do not change their sign across the minor axis while rotational motions do change their sign.

We now turn our attention to velocity channel maps. Representative velocity channels for different inclination are shown in Figure \ref{fig:channel3_2mjup}. Channel maps for standard $0.5~M_{\rm Jup}$ and $1~M_{\rm Jup}$ models are presented in Appendix \ref{sec:channel_maps}. While the characteristic features of buoyancy spirals are still present, they are clearly weaker for inclined disks due to the $\sin i$ components of the projection, in an agreement with what we see in Keplerian-subtracted moment maps. This suggests that disks with small inclination of $\lesssim 30^\circ$ offer the best opportunities to search for buoyancy spirals.

\section{DISCUSSION}
\label{sec:discussion}

\begin{figure*}[]
    \centering
    \includegraphics[width=\textwidth]{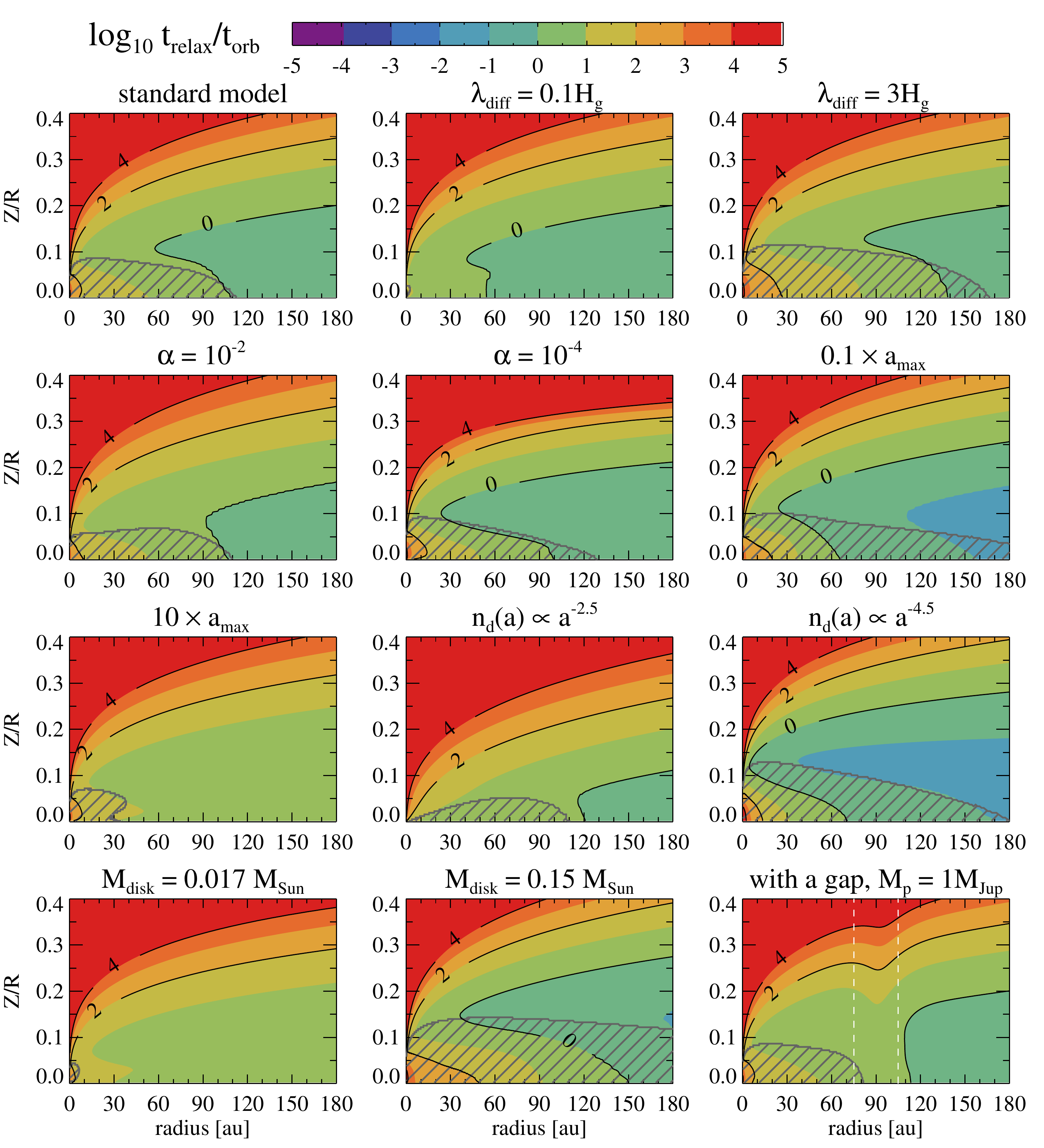}
    \caption{Same as Figure \ref{fig:cooling_time}f, which is shown again in the upper left panel, but the relaxation timescale calculated with various parameters. Unless otherwise stated in the title of each panel, the fiducial parameters adopted to compute the diffusion and collisional timescales are $\lambda_{\rm diff} = H_{\rm g}$, $\alpha=10^{-3}$, $a_{\rm max} (R) = 1~{\rm mm}~(R/30~{\rm au})^{-2}$, and $n_{\rm d} (a) \propto a^{-3.5}$. The shaded region in gray shows where $t_{\rm diff} \geq t_{\rm coll}$. In the bottom right panel, the two dashed lines show the radial locations at which the gap depth is the half of the maximum (75 and 105~au). We note that, for the broad range of parameter space explored, the thermal relaxation of the gas in the surface layers of protoplanetary disks (i.e., $Z/R \gtrsim 0.1$) is limited due to infrequent gas-grain collisions.}
    \label{fig:tcool_zr_param}
\end{figure*}

\subsection{Thermal Relaxation Timescale}
\label{sec:t_relax}

As we have shown with hydrodynamic simulations in Section \ref{sec:results}, the development of buoyancy resonances depends on the local thermal relaxation timescale. To examine the development and observability of buoyancy resonances under various disk conditions, we explore a broad range of parameter space and compute the relaxation timescale following the approach described in Section \ref{sec:thermal_relaxation}. Specifically, we vary (1) the diffusion length scale $\lambda_{\rm diff}$, (2) the dust scale height, (3) the maximum dust grain size, (4) the power-law slope in the dust size distribution, and (5) the disk mass. The resulting thermal relaxation timescales are shown in Figure \ref{fig:tcool_zr_param}. We compare $t_{\rm relax}$ with $N_Z^{-1}$ in Appendix \ref{sec:t_relax_Nz} and Figure \ref{fig:tcoolN}. In what follows, we focus our discussion on the surface layers of the disk ($Z/R \gtrsim 0.1$) as we are mostly concerned about the observability of buoyancy resonances using optically thick CO lines which will trace elevated regions of the disk. 

Due to the steep dependency ($t_{\rm diff} \propto \lambda^2_{\rm diff}$), having a short diffusion length scale of $\lambda = 0.1~H_{\rm g}$ results in a diffusion timescale that is much shorter than the dynamical timescale almost everywhere in the disk. Even in such a case, however, the relaxation of the gas is limited by long collisional timescales between gas molecules and dust grains. In particular, the relaxation timescale in the surface layers where $^{12}$CO lines probe is determined by the collisional timescale and is insensitive to the choice of $\lambda_{\rm diff}$.

The gas-dust collision rate is dependent upon the detailed spatial and size distribution of grains. We first vary the scale height of dust grains by changing $\alpha$ in Equation (\ref{eqn:scale_dust}). This has two opposite effects on the collisional timescale. For an increased scale height, the mean dust size increases which would lengthen the collisional timescale (Equation \ref{eqn:tcool_coll2}). At the same time, the dust-to-gas mass ratio also increases which would shorten the collisional timescale. The two effects effectively cancel out and the collisional timescale in the surface layers remains sufficiently long for buoyancy resonances to develop, as can be seen in Figures \ref{fig:tcool_zr_param} and \ref{fig:tcoolN}. 

When $a_{\rm max}$ is decreased, the mean grain size decreases. The dust-to-gas mass ratio increases in the surface layers because, with a smaller $a_{\rm max}$, a larger fraction of the total dust mass is in the grains that can be lofted sufficiently high. Together, this shortens the collisional timescale, resulting in $t_{\rm relax} < N_Z^{-1}$ in a larger region of the disk as shown in Figure \ref{fig:tcoolN}. Nevertheless, $t_{\rm relax} \gtrsim N_Z^{-1}$ near the $^{12}$CO emission surface, suggesting that buoyancy resonances likely develop there. 

Changes in the power-law index of the dust size distribution work in a similar way. When the dust size distribution follows a steeper power-law distribution, a larger fraction of the total dust mass is in small grains that can be lofted up high. This will result in a smaller mean grain size and a larger dust-to-gas mass ratio, which would shorten the collisional timescale. 

Next, we vary the disk mass by a factor of three. Having a larger disk mass increases the number of colliders, increasing the gas-dust collision rate. The relaxation timescale would be therefore shortened, while $t_{\rm relax } \gtrsim N_Z^{-1}$ in the majority of the disk regions.

Lastly, we adopt the azimuthally-averaged gas density from the standard $1~M_{\rm Jup}$ model to examine the influence the gap has on the relaxation timescale. Because dust grains are depleted within the gap the collisional timescale becomes longer, facilitating the development of buoyancy resonances there.

In summary, we argue that the relaxation timescale of the gas in the surface layers ($Z/R \gtrsim 0.1$) is limited by infrequent gas-dust collisions under a broad range of conditions applicable to protoplanetary disks (Figure \ref{fig:tcool_zr_param}). This is because the dust-to-gas mass ratio is small ($\rho_{\rm d}/\rho_{\rm g} \lesssim 10^{-3}$) in the surface layers due to the vertical settling of dust grains. The resulting thermal relaxation timescale is comparable to or longer than the timescale associated with buoyancy oscillations $N_Z^{-1}$ (Figure \ref{fig:tcoolN}), suggesting that the surface layers have a favorable condition for buoyancy resonances to develop.

\subsubsection{Implications for planet-induced gaps}

While we focused our analysis mainly on the surface layers so far, it is worth pointing out the finite relaxation timescale near the disk midplane. Even at large radii for which it is typically thought that less absorbing materials would result in more efficient cooling, we find that the infrequent gas-dust collision could prevent the gas from cooling instantaneously. 

The finite relaxation timescale in the main body of a disk can have important implications for the formation of gaps by planets. Adopting a constant dimensionless relaxation timescale $\beta \equiv 2 \pi t_{\rm relax}/t_{\rm orb}$ in vertically-integrated two-dimensional disks, \citet{miranda20a} and \citet{zhang20} independently showed that the isothermal assumption does not provide a good approximation when $\beta \gtrsim 10^{-3}$ -- note again that we consistently find $\beta \gtrsim 10^{-3}$ under various disk conditions (see Figure \ref{fig:tcool_zr_param}). In particular, when $\beta \sim 1$ the radiative dissipation of Lindblad spirals becomes important and the gap around the planet becomes narrower than otherwise. 

\begin{figure}[t]
    \centering
    \includegraphics[width=0.48\textwidth]{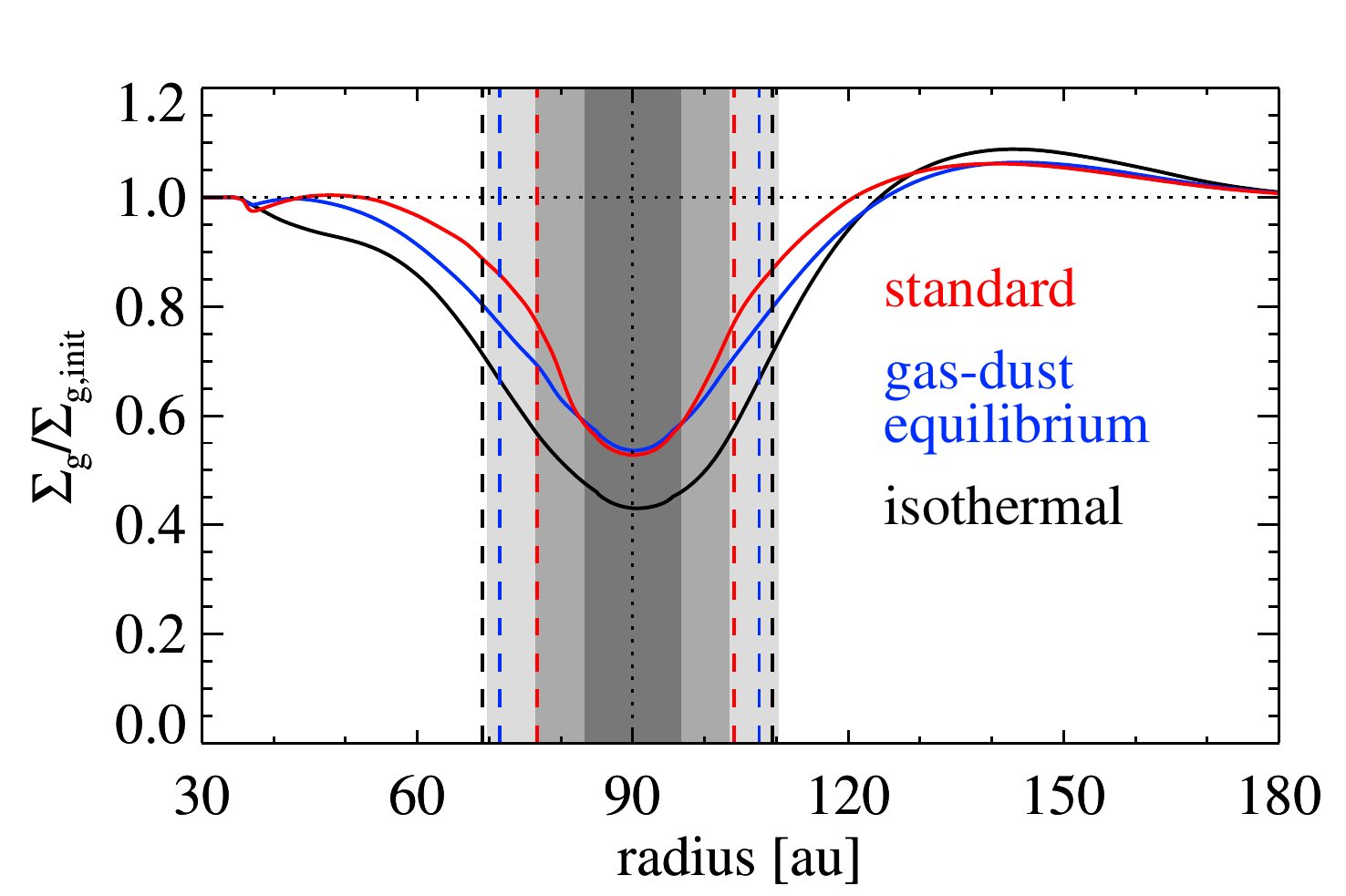}
    \caption{The azimuthally-averaged gas surface density as a function of radius. The hill sphere is excluded when azimuthally averaging the density. Red, blue, black curves show the profiles from the standard model, the gas-dust thermal equilibrium model, and the isothermal model, respectively. The vertical dashed lines present the gap width at the half maximum. The gray shaded regions show $\pm$1, $\pm$2, and $\pm$3 scale heights from the planet.}
    \label{fig:gap}
\end{figure}

In Figure \ref{fig:gap}, we present the radial profiles of the gas surface density from the standard model and the gas-dust equilibrium model. We additionally ran a simulation adopting an isothermal equation of state and included the density profile from the isothermal simulation in the same figure. The gap widths measured at the half maximum are $\Delta_{\rm gap} = $ 27.4~au, 36.1~au, and 40.4~au for the three models. Note that the gap is wider in the isothermal simulation, consistent with previous two-dimensional simulations \citep{miranda20a,zhang20}. 

There have been many attempts to infer masses of planets responsible for observed gaps, using planet-disk interaction simulations or empirical planet mass -- gap width relations (see \citealt{bae18,diskdynamics20} and references therein). We note that most of the simulations were carried out adopting an isothermal equation of state, implying that the inferred planet masses can underestimate the actual planet masses. Along the same lines, we argue that empirical gap depth/width -- planet mass relations that are typically derived under the isothermal assumption need a revision.

\begin{figure*}[]
    \centering
    \includegraphics[width=1.05\textwidth]{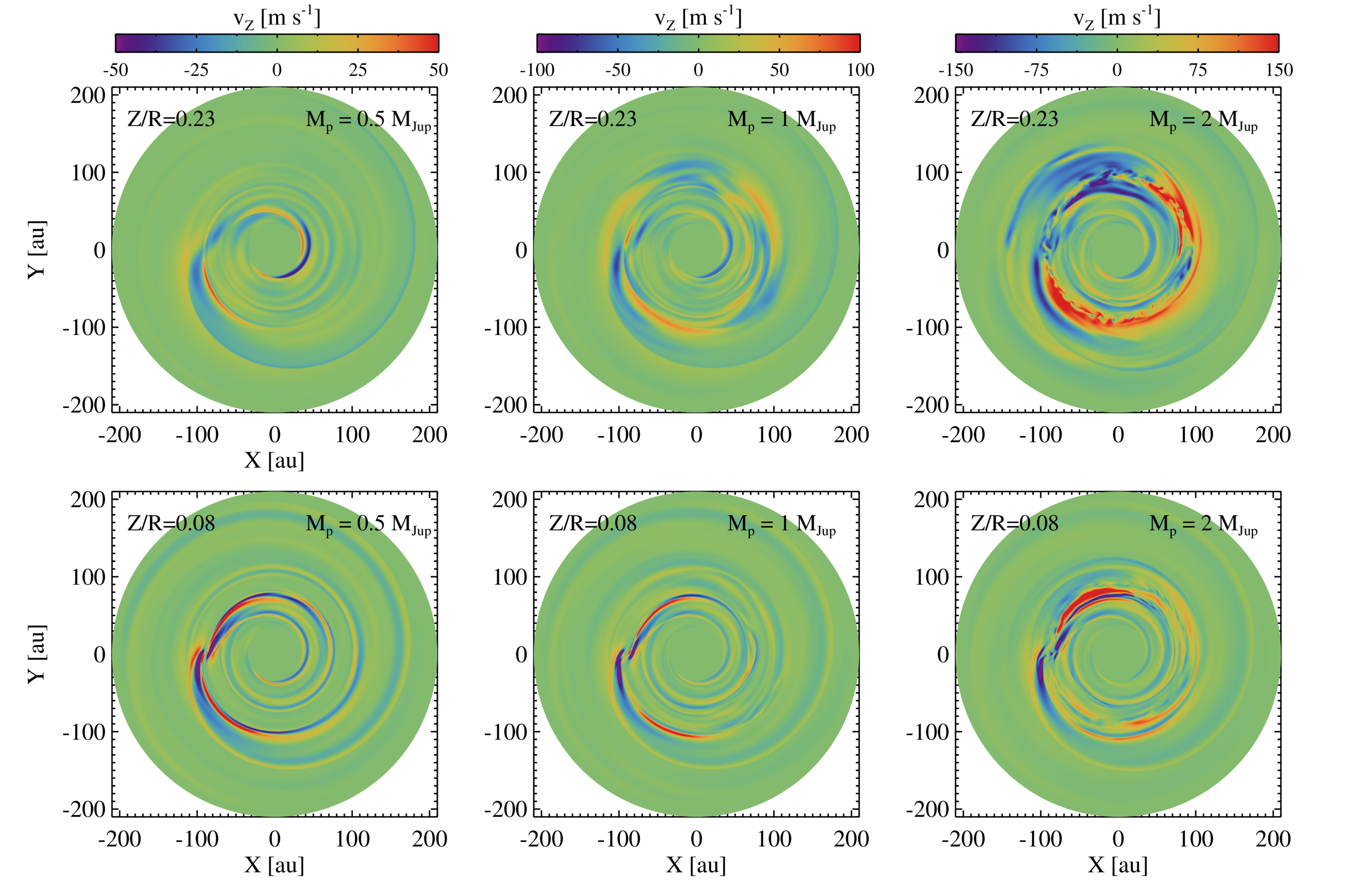}
    \caption{Same as Figure \ref{fig:hydroxy} but for models with rapid cooling near the surface using $Z_{\rm line} = 2H_{\rm g}~(Z/R \simeq 0.15)$.}
    \label{fig:hydroxy_zline2h}
\end{figure*}
\begin{figure*}[]
    \centering
    \includegraphics[width=\textwidth]{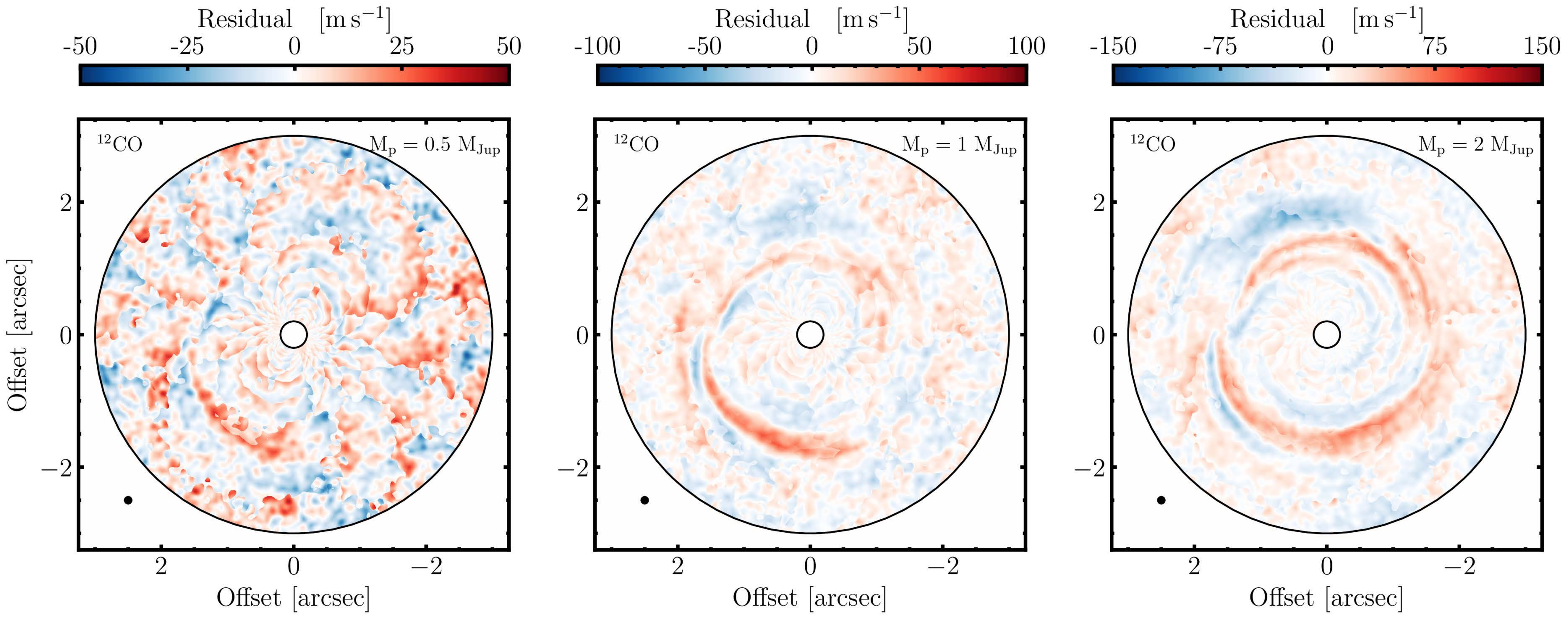}
        \includegraphics[width=\textwidth]{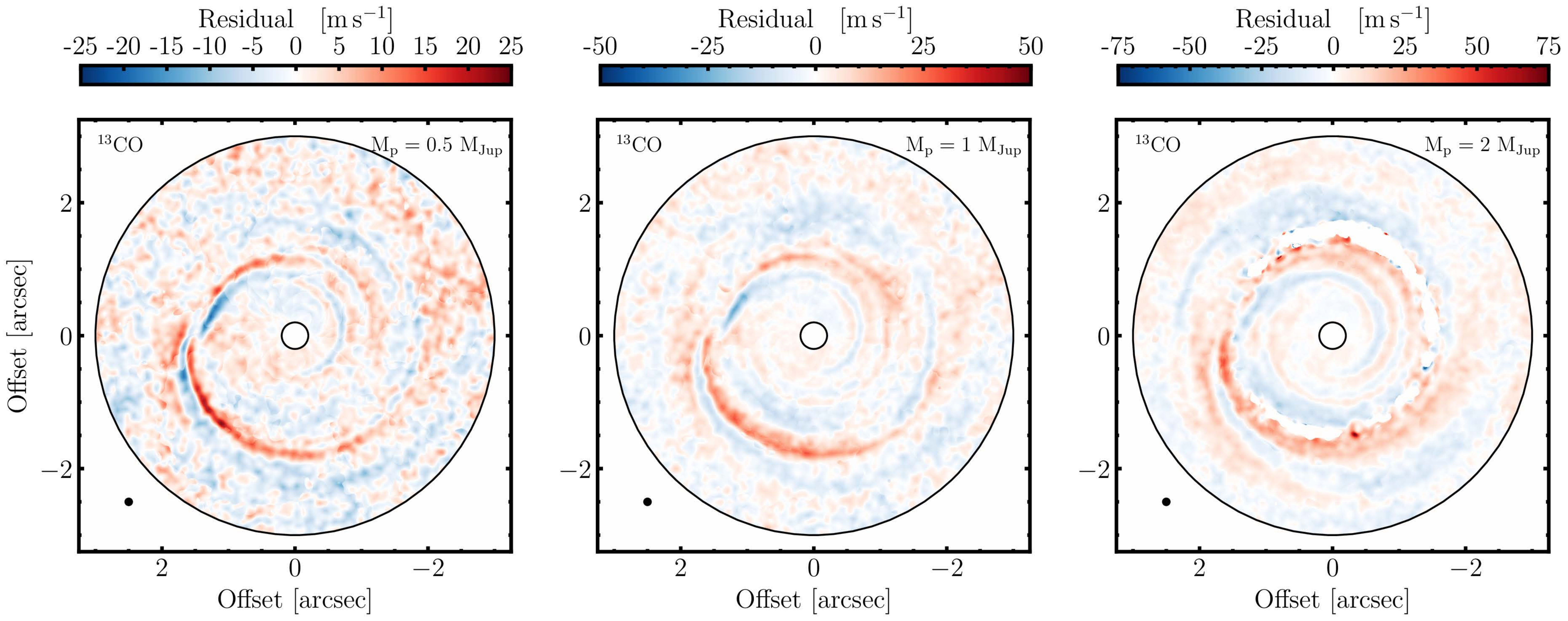}
    \caption{Same as Figure \ref{fig:simobs} but for models with rapid cooling near the surface using $Z_{\rm line} = 2H_{\rm g}$.}
    \label{fig:simobs_zline2h}
\end{figure*}

\subsubsection{Implications for hydrodynamic instabilities}

Hydrodynamic instabilities are an important source of turbulence in protoplanetary disks. Vertical shear instability can be largely suppressed with finite thermal relaxation timescales \citep{nelson13,lin15,malygin17,pfeil19,pfeil20}. Spiral wave instability is known to have a forbidden region near the disk surface where the buoyancy frequency is larger than a half of the Doppler-shifted forcing frequency \citep{bae16b}. The forbidden region extends toward the midplane with more adiabatic gas response \citep{bae16b}, so we can infer that spiral wave instability operates in a more confined region with finite thermal relaxation timescales. On the other hand, finite thermal relaxation timescales can help other instabilities that operate with slower thermal relaxation, such as convective over stability \citep{klahr14} and zombie vortex instability \citep{marcus15,barranco18}.

\subsection{Effect of Efficient Line Cooling Near the Surface}
\label{sec:zline_2h}

As discussed in Section \ref{sec:thermodynamics}, exactly at which height atomic/molecular line cooling becomes the dominant cooling mechanism depends upon the balance between various heating and cooling mechanisms, which in turn is determined by various factors including the amount of dust grains, gas temperature, abundance of coolant atoms/molecules, and electron number density. Here, we test the effect of efficient line cooling near the surface, by adopting $Z_{\rm line} = 2~H_{\rm g}$ in Equation (\ref{eqn:t_relax}) (c.f., $Z_{\rm line} = 4~H_{\rm g}$ in the standard model).

Figure \ref{fig:hydroxy_zline2h} presents the vertical velocity distribution. Not surprisingly, buoyancy resonances  are weaker at $Z = 3H_{\rm g}~(Z/R = 0.23)$ due to the rapid cooling in the surface layers. However, we note that buoyancy resonances are not completely suppressed. This is because even when cooling is rapid (effectively $\gamma \simeq 1$), the buoyancy frequency is not strictly zero due to the stratified temperature profile (see Equation \ref{eqn:buoyancy_frequency_ideal}). At $Z = 1 H_{\rm g}~(Z/R = 0.08)$, buoyancy resonances develop at the level they develop in the fiducial model adopting $Z_{\rm line} = 4H_{\rm g}$ (see Figure \ref{fig:hydroxy}). This suggests that the development of buoyancy resonances depends on the {\it local} thermodynamic properties and that, as far as line cooling is not efficient all the way to the midplane, there will be regions where buoyancy resonances would develop. 

From the observational point of view, this suggests that we can choose optically thiner lines that probe the adequate heights. To support this argument, we generate Keplerian-subtracted moment maps from the models with efficient line cooling, which are shown in Figure \ref{fig:simobs_zline2h}. The residual velocities are smaller than the fiducial model in $^{12}$CO because the line probes the surface layers where cooling is rapid. On the other hand, the morphology and magnitude of the buoyancy spirals in the $^{13}$CO maps are nearly identical to the standard model (see Figure \ref{fig:simobs}).

\subsection{Can we observationally distinguish the origin of non-Keplerian motions?}

Here we discuss potential ways to discriminate the non-Keplerian motions driven by buoyancy resonances from those driven by other origins, Lindblad resonance, corrugated vertical flows, and gas pressure changes.

{\it Lindblad spirals:}
The vertical dependency of the perturbation driven by buoyancy and Lindblad spirals is the key to discriminate the two. Because buoyancy frequency is strictly zero at the disk midplane, we expect no or weak buoyancy resonances there. On the other hand, perturbations driven by Lindblad spirals are the strongest at the disk midplane and decrease over height (Figures \ref{fig:vcomp} and \ref{fig:azimuth_wtcoll}). Observations of multiple lines tracing different heights in the disk, for instance $^{12}$CO vs. $^{13}$CO or C$^{18}$O, will thus help discriminate between buoyancy spirals and Lindblad spirals.

\begin{figure*}[ht!]
    \centering
    \includegraphics[width=\textwidth]{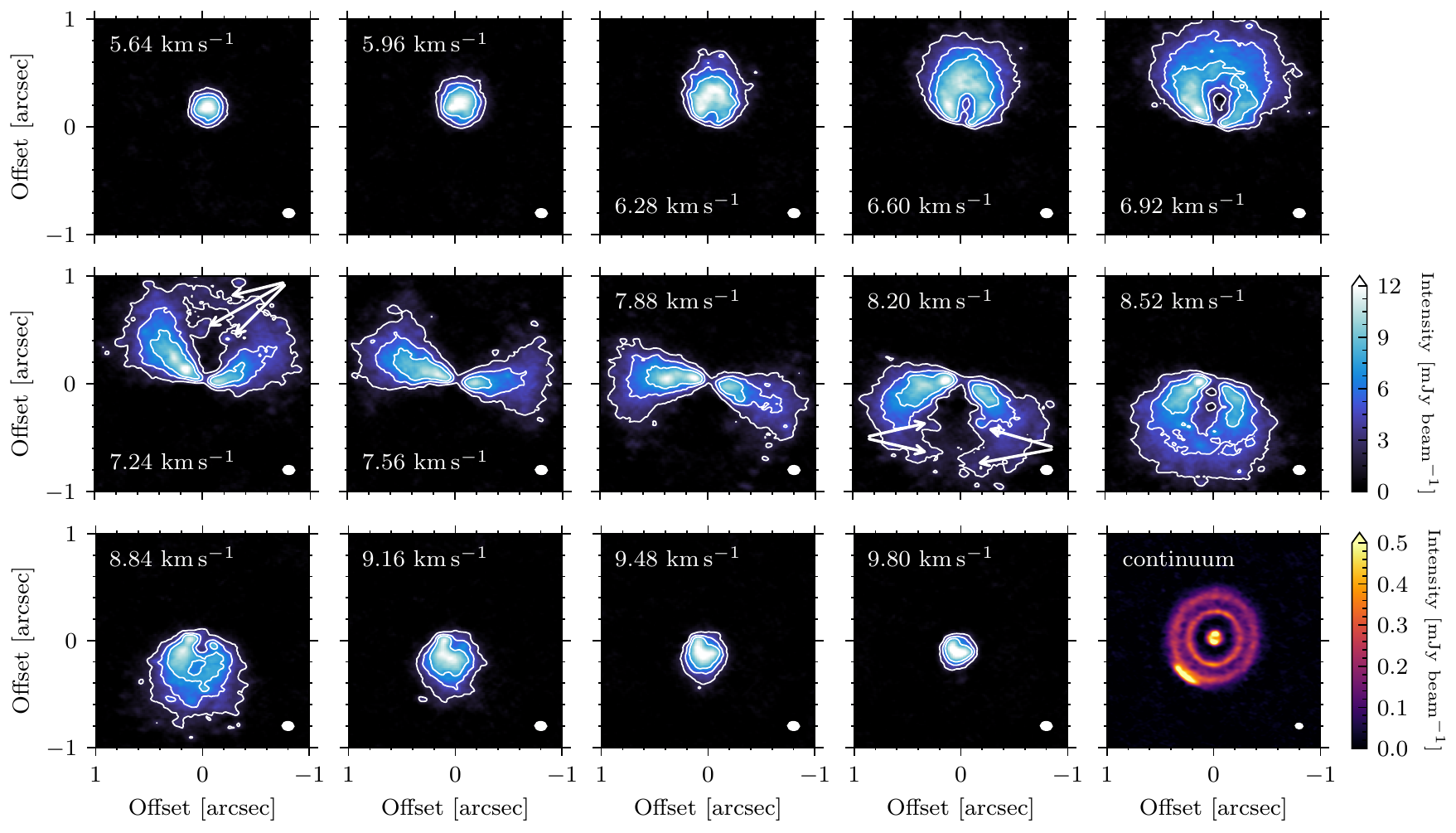}
    \caption{$^{12}$CO channel maps of HD~143006, originally presented in \citet{lperez18}. Contours are drawn at 2.4, 4.8, and $7.2~{\rm mJy~beam}^{-1}$. The rms noise is $0.3~{\rm mJy~beam}^{-1}$. The bottom right panel shows the continuum emission of the disk. Note the CO arcs connecting the Keplerian wings at $\sim 0.4''$ and $0.8''$ from the center, most clearly seen on the North side of the disk at $7.24~{\rm km~s}^{-1}$ and on the South side of the disk at $8.20~{\rm km~s}^{-1}$. These non-Keplerian velocity components resemble what we expect from buoyancy spirals.}
    \label{fig:hd143006}
\end{figure*}

{\it Corrugated vertical flows:}
Various (magneto-)hydrodynamic processes are known to create radially-alternating corrugated vertical flow patterns, including vertical shear instability \citep{nelson13}, spiral wave instability \citep{bae16b,bae16a}, and magnetically-driven zonal flows \citep{johansen09,flock15,riols20}. Vertical velocity perturbations driven by buoyancy resonances (and Lindblad resonance) are symmetric against the midplane, so at a given radius the gas motion will be either toward the midplane or toward the surface. In contrast, vertical velocity perturbations associated with corrugated vertical flows are not symmetric against the midplane. Rather, instabilities develop into corrugation modes and the entire column oscillates vertically \citep{nelson13,bae16b,bae16a}. This difference suggests that we can distinguish the two scenarios if we probe velocity perturbations at upper and lower surfaces of the disk separately. Optically thick tracers (e.g., $^{12}$CO), a sufficiently high spatial resolution, and a moderately inclined disk geometry would provide the best chance to separate the upper and lower surface emission.

{\it Gas pressure changes:}
Gas pressure changes across radius can lead sub-/super-Keplerian rotation of the gas to maintain the radial force balance \citep{teague18}. As we discussed in Section \ref{sec:inclination}, distinguishing non-Keplerian motions associated with buoyancy spirals from rotation modulation can become a challenge in inclined disks as we become sensitive to both vertical and azimuthal velocities. As an example, we present channel maps of $^{12}$CO line emission of HD~143006 in Figure \ref{fig:hd143006}. Morphologically, the arcs connecting the Keplerian wings (most clearly seen on the North side of the disk at $v=7.24~{\rm km~s}^{-1}$ and on the South side of the disk at $v=8.20~{\rm km~s}^{-1}$) show a good resemblance to those features expected from buoyancy resonance. However, these arcs can instead be interpreted as rotation modulation. The non-Keplerian features are most prominently seen as red-shifted arcs in blue-shifted channels (e.g., $v=7.24~{\rm km~s}^{-1}$) and blue-shifted arcs in red-shifted channels (e.g., $v=8.20~{\rm km~s}^{-1}$), which are consistent with the expected modulation associated with sub-Keplerian rotation \citep{teague19}. In fact, the inner arcs at $\sim0.4''$ coincide with the outermost continuum ring, suggesting that the arcs could arise from the rotation modulation around the pressure peak. The outer arcs at $\sim0.8''$ is close to where $^{12}$CO emission fades, suggesting they could be due to a rapid drop in the gas density at that radius. Because of the degeneracy between vertical and azimuthal velocities in channel maps, we recommend to use Keplerian-subtracted moment maps rather than velocity channel maps when it comes to distinguishing buoyancy spirals and rotation modulation arising from gas pressure changes (see Section \ref{sec:inclination}).

\begin{figure*}[ht!]
    \centering
    \includegraphics[width=\textwidth]{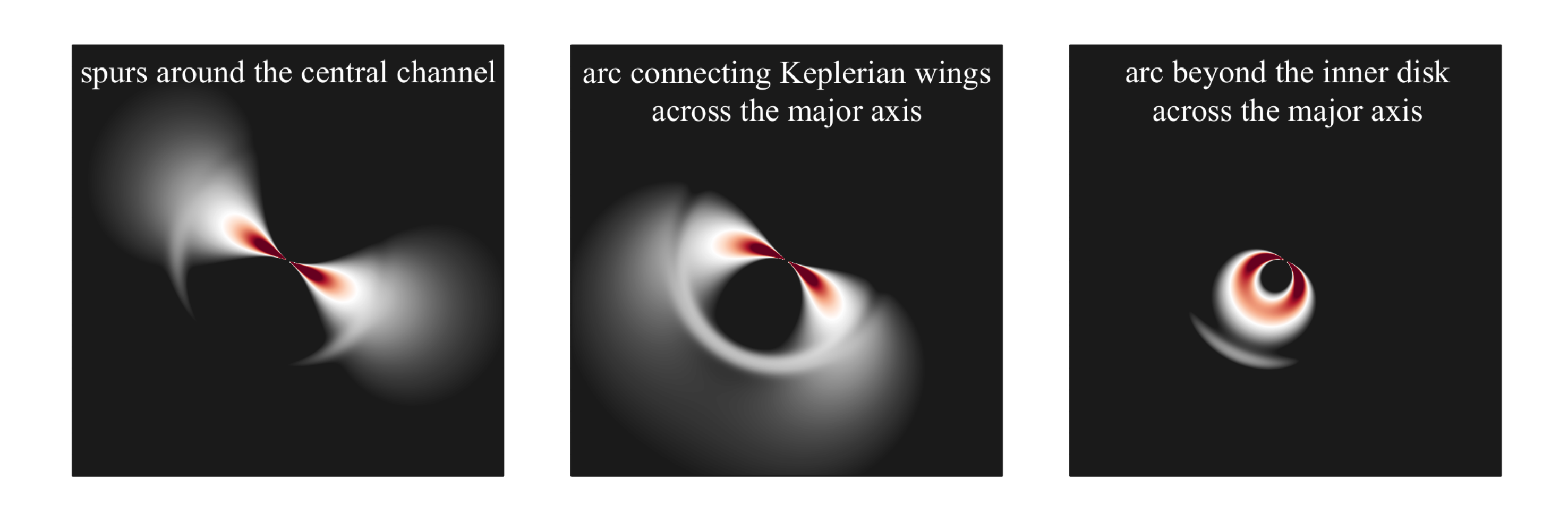}
    \caption{Cartoon channel maps summarizing main features expected from buoyancy spirals: (left) spurs around the central velocity channel, standing out of the Keplerian wings; (middle) an arc connecting Keplerian wings across the semi-major axis of the disk; and (right) an arc beyond the inner disk across the semi-major axis of the disk.     }
    \label{fig:channel_cartoon}
\end{figure*}

\subsection{Buoyancy Resonances and Dust}

How would buoyancy resonances appear in scattered light and (sub-)millimeter continuum observations? Interestingly enough, \citet{vanboekel17} reported a tightly wound spiral in the near-infrared polarized intensity map of the TW~Hya disk. The spiral is located at the outer edge of an annular gap centered at about 93~au\footnote{This number is updated from \citet{vanboekel17} in accordance with the {\it Gaia} distance of 60.1~pc \citep{bailer-jones18}.}, and extends about 90 degrees in azimuth on the South-West side of the disk. While we opt out of making simulated scattered light observations from our models because our simulations do not include dust grains, it is interesting to point out that a buoyancy spiral produces positive density perturbations at the exact location where the scattered light spiral is revealed (see the lower-right quadrant of $\delta\rho/\langle \rho\rangle_\phi$ at $Z/R=0.08$ and $0.23$ in Figure \ref{fig:vcomp}). It is interesting to speculate that the velocity spiral in the CO observation and the density spiral in the scattered light observation are probing buoyancy resonances driven by a planet embedded within the gap at $\sim90$~au. Future simulations including dust particles will help further investigate this possibility.

On the other hand, given that buoyancy resonances are weak or absent near the midplane and that buoyancy spirals are confined in the corotating region where large (sub-)millimeter-sized grains are expected to depleted due to radial drift, we believe observing buoyancy resonances in (sub-)millimeter continuum observations is less likely. 

\section{SUMMARY AND CONCLUSION}
\label{sec:summary}

Along with the spirals driven by the well-known Lindblad resonance, we showed that planets can excite spirals via buoyancy resonances, which we can detect using molecular line observations. We summarize our findings below.

(1) Under a broad range of conditions applicable to protoplanetary disks, we showed that infrequent gas-dust collision can be the bottleneck in the energy exchange between the gas and dust in the surface layers ($Z/R \gtrsim 0.1$; Figures \ref{fig:cooling_time} and \ref{fig:tcool_zr_param}). Although this has been previously suggested  \citep{malygin17,barranco18,pfeil19}, to our knowledge, it is the first time that this effect is taken into account in planet-disk interaction simulations.

(2) The collision-limited slow thermal relaxation provides favorable conditions for buoyancy resonances to develop (Figures \ref{fig:t_relax_Nz} and \ref{fig:tcoolN}). Adopting the thermal relaxation timescale estimated by considering radiation, diffusion, and gas-dust collision, we showed that planets can excite a family of tightly wound spirals via buoyancy resonances, in addition to those excited by Lindblad resonance (Figures \ref{fig:hydro_wtcoll} and \ref{fig:vcomp}). 

(3) Two main characteristics of buoyancy spirals are their small pitch angles and large vertical motions. Buoyancy spirals have a pitch angle of a few to 10~degrees in the corotating region of the planet (Figure \ref{fig:pitchangle}). The vertical motions associated with buoyancy resonances is of order of $100~{\rm m~s}^{-1}$ for Jovian-mass planets, corresponding to about $20~\%$ of the sound speed or a few $\%$ of the Keplerian speed. This is comparable to or larger than the velocity perturbations driven by Lindblad resonance (Figures \ref{fig:vcomp} and \ref{fig:azimuth_wtcoll}).

(4) By generating synthetic ALMA observations, we showed that the non-Keplerian motions associated with buoyancy resonances is detectable. Buoyancy spirals would appear as tightly-wound arcs in Keplerian-subtracted moment maps (Figure \ref{fig:simobs}). In velocity channel maps, buoyancy spirals appear as spurs around the central velocity channel, arcs connecting Keplerian wings or an arc beyond the inner disk across the semi-major axis of the disk. We summarize these features with a cartoon in Figure \ref{fig:channel_cartoon}.

(5) Because buoyancy resonances predominantly produce vertical velocity perturbations, face-on disks provide the best opportunities to search for their signatures. Based on the morphology and the magnitude of velocity perturbations, we propose that the tightly wound spirals seen in the near-face-on TW Hya disk \citep{teague19} could be driven by a (sub-)Jovian-mass planet at $\sim90$~au. 

(6) Along with the implications for buoyancy resonances, the finite relaxation timescale has important implications for planet-induced gaps and. As shown in \citet{miranda20a} and \citet{zhang20}, when the cooling of the disk gas is moderate (i.e., $\beta \equiv 2\pi t_{\rm relax}/t_{\rm orb} \sim 1$) the gap around the planet becomes narrower than that in fully isothermal ($\beta \ll 1$) or fully adiabatic ($\beta \gg 1$) simulations. The finite relaxation timescale implies that the mass of planets responsible for gaps seen in continuum observations can be underestimated if inferred based on isothermal simulations. 

(7) We discussed potential ways to distinguish non-Keplerian motions driven by buoyancy resonances and those driven by other mechanisms: Lindblad resonance, corrugated vertical flows, and gas pressure changes. We recommend the community to observe multiple lines tracing different heights in the disk. It is also crucial to have sufficiently high spatial/spectral resolution and sensitivity to separate the emission arising from the near and far sides of the disk.

We conclude by emphasizing that numerical simulations have to include more realistic and complete treatments for thermodynamics to fully capture planet-disk interaction. Planet-disk interaction simulations often (but not always) adopt a vertically isothermal temperature structure and/or an isothermal equation of state. When it comes to buoyancy resonances, such simplified models can completely suppress the resonance. We should point out that our simulations have caveats. We adopted a prescribed, fixed thermal relaxation model. In reality, the spatial and size distribution of dust grains would evolve over time, and this is neglected in current simulations. It will be also interesting to implement a thermo-chemistry model that evolves over time, coupled with the hydro evolution. Future simulations with more complete treatments for dust- and thermo-dynamics will help better interpret state-of-the-art observations.

\acknowledgments

We thank the anonymous referee for providing us a helpful report that improved the initial manuscript. JB is thankful to Myriam Benisty, Stefano Facchini, and Steve Lubow for helpful conversations. JB acknowledges support by NASA through the NASA Hubble Fellowship grant \#HST-HF2-51427.001-A awarded  by  the  Space  Telescope  Science  Institute,  which  is  operated  by  the  Association  of  Universities  for  Research  in  Astronomy, Incorporated, under NASA contract NAS5-26555. JB acknowledges the computational resources and services provided by the Extreme Science and Engineering Discovery Environment (XSEDE), which is supported by National Science Foundation grant number ACI-1548562, and by the NASA High-End Computing (HEC) Program through the NASA Advanced Supercomputing (NAS) Division at Ames Research Center. RT acknowledges support from the Smithsonian Institution as a Submillimeter Array (SMA) Fellow. ZZ acknowledges support from the National Science Foundation under CAREER Grant Number AST-1753168.

\software{\texttt{FARGO3D} \citep{benitez16}, \texttt{RADMC-3D} \citep{radmc3d}, \texttt{eddy} \citep{eddy}}

\appendix

%
%

\section{The Relaxation Timescale}
\label{sec:t_relax_Nz}

\begin{figure*}[h!]
    \centering
    \includegraphics[width=1\textwidth]{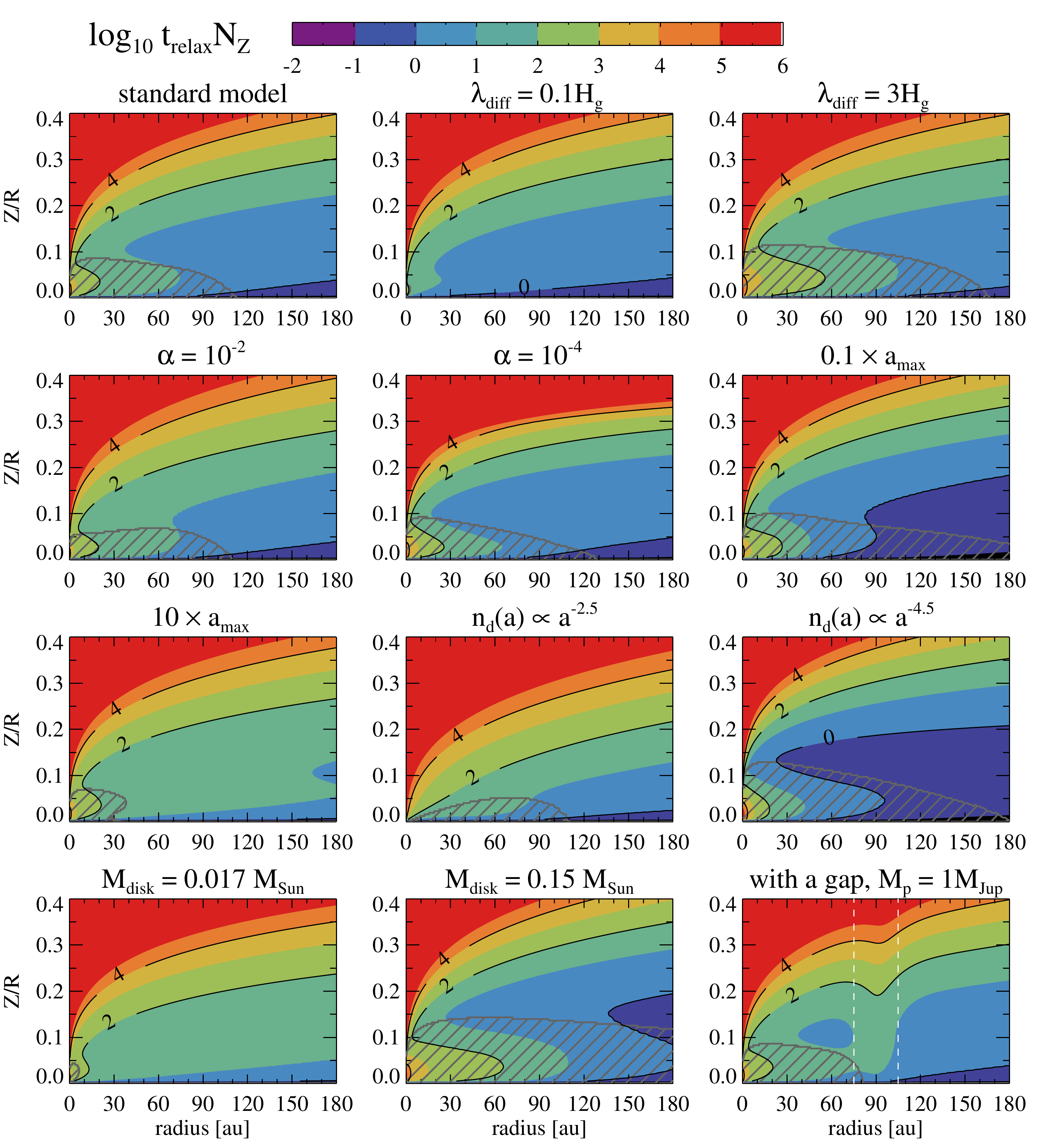}
    \caption{Similar to Figure \ref{fig:tcool_zr_param}, but showing $t_{\rm relax} N_Z$.}
    \label{fig:tcoolN}
\end{figure*}

In Figure \ref{fig:tcoolN} we present $t_{\rm relax} N_Z$ for the disk models discussed in Section \ref{sec:t_relax}. For the broad range of parameter space we explored, $t_{\rm relax} \gtrsim N_Z^{-1}$ in $Z/R \gtrsim 0.1$, suggesting that buoyancy resonances likely develop in the surface layers of protoplanetary disks.

\section{Additional Channel Maps}
\label{sec:channel_maps}

\begin{figure*}[]
    \centering
    \includegraphics[width=0.95\textwidth]{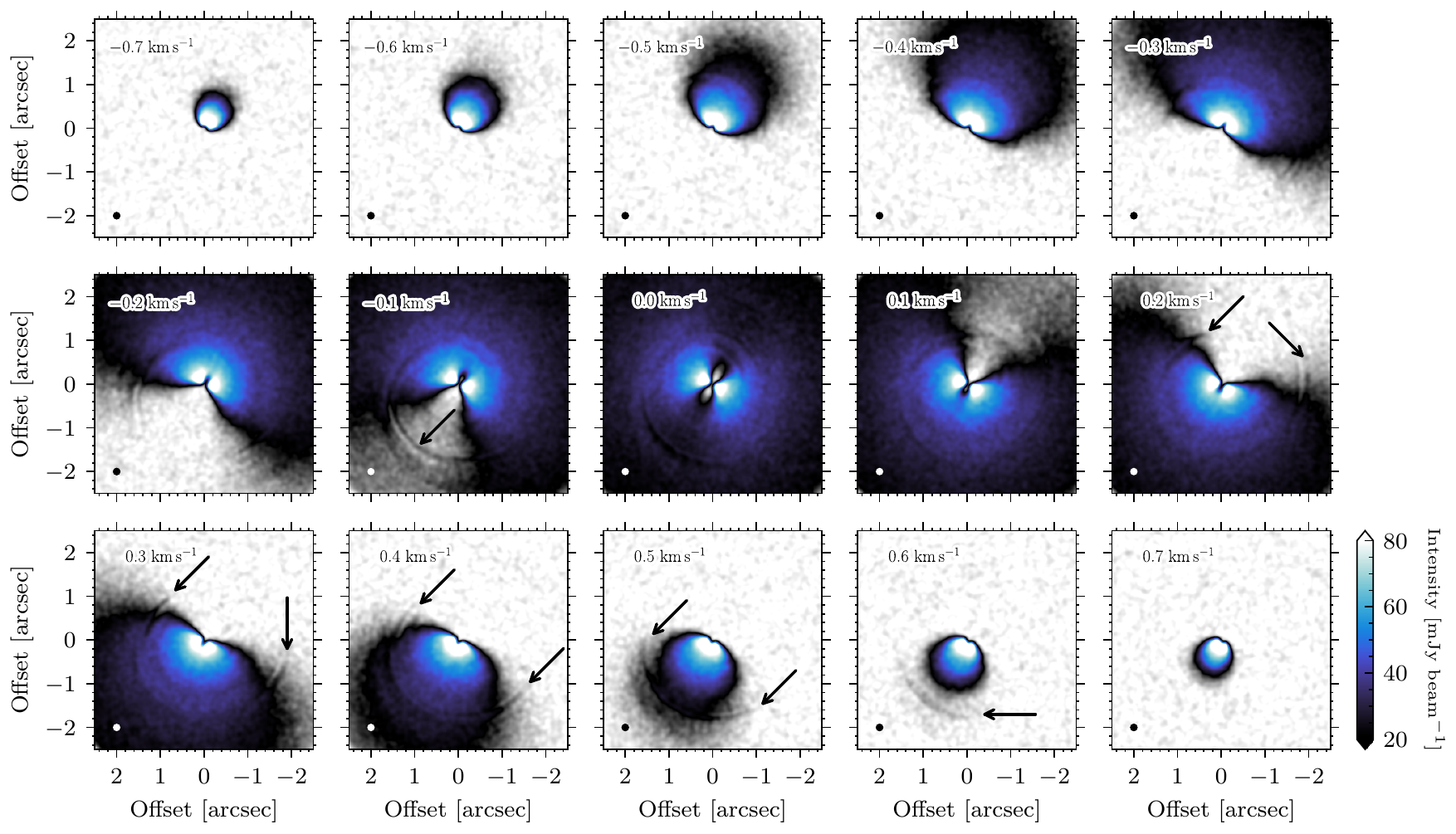}
    \caption{Same as Figure \ref{fig:channel3_2mjup}, but with $M_p = 0.5~M_{\rm Jup}$.}
    \label{fig:channel3_05mjup}
\end{figure*}

\begin{figure*}[h!]
    \centering
    \includegraphics[width=0.95\textwidth]{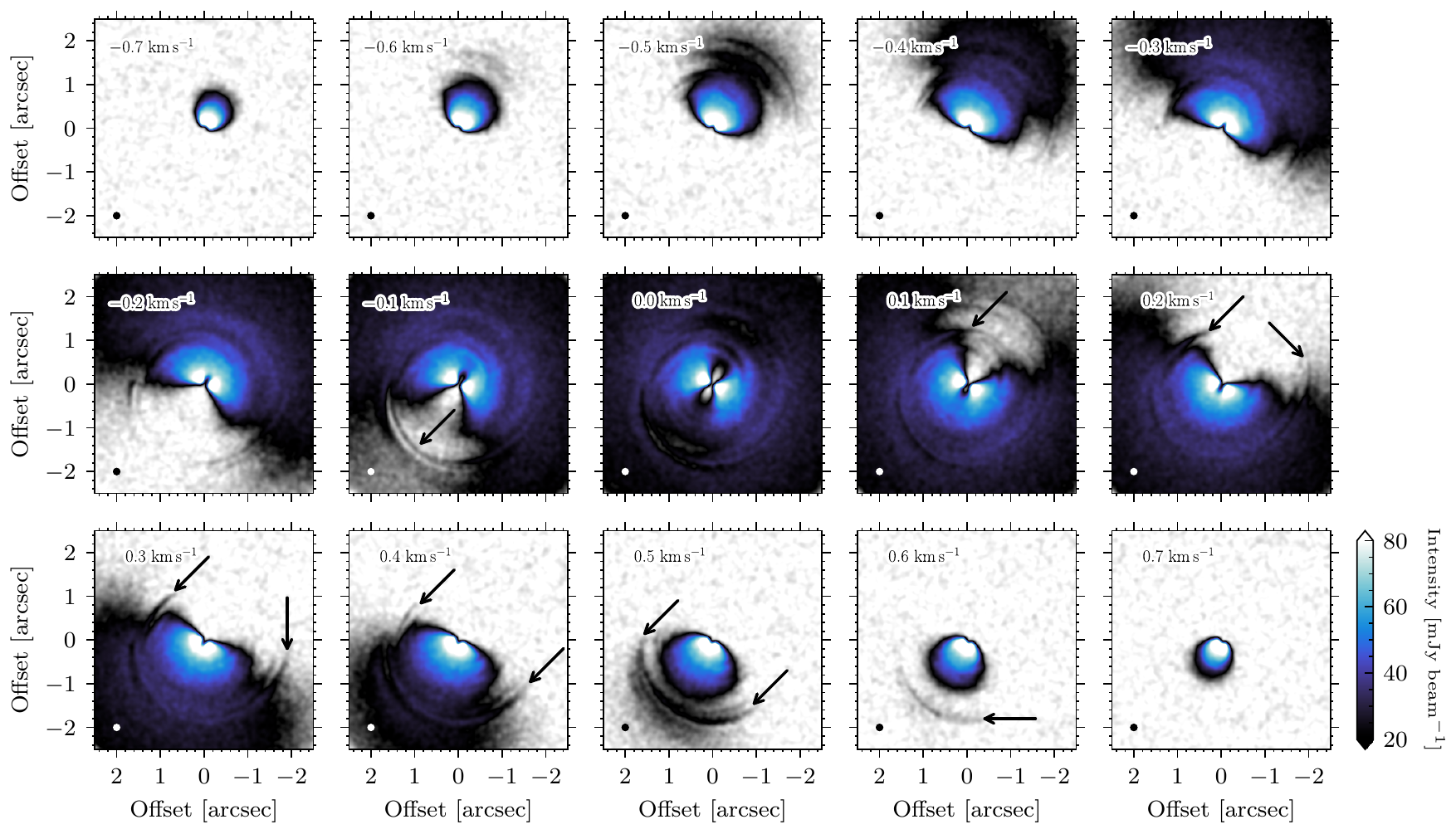}
    \caption{Same as Figure \ref{fig:channel3_2mjup}, but with $M_p = 1~M_{\rm Jup}$.}
    \label{fig:channel3_1mjup}
\end{figure*}

Figures \ref{fig:channel3_05mjup} and \ref{fig:channel3_1mjup} present channel maps from synthetic $^{12}$CO observations of models with $0.5~M_{\rm Jup}$ and $1~M_{\rm Jup}$ planets. Simulated cubes are publicly available at \url{https://doi.org/10.5281/zenodo.4361639}.


\bibliography{bibliography}{}
\bibliographystyle{aasjournal}

\end{document}